\documentclass[pdftex,pagesize=auto,paper=letter,11pt,abstract=on,bibliography=oldstyle]{scrartcl}\usepackage{wrapfig}\usepackage[pdftex]{xcolor}\usepackage[pdftex]{graphicx}\usepackage{tikz,pgfplots}\usepackage[utf8]{inputenc}\usepackage{amssymb,amsmath}\usepackage{url}\usepackage{xspace}\usepackage{rotating}\usepackage{booktabs}\usepackage[bookmarks=false]{hyperref}\usepackage{relsize}\usepackage[sort,compress]{cite}\usepackage[format=plain,labelformat={simple},font={small},labelsep={period},labelfont={bf}]{caption}\usepackage{authblk}\usepackage[T1]{fontenc}\usepackage{lmodern}\usepackage{textcomp}\usepackage{microtype}\def\comment#1{}\def\withcomments{\newcounter{mycommentcounter}\def\comment##1{\refstepcounter{mycommentcounter}\ifhmode\unskip{\dimen1=\baselineskip \divide\dimen1 by 2 \raise\dimen1\llap{\tiny\bfseries \textcolor{red}{-\themycommentcounter-}}}\fi\marginpar[{\renewcommand{\baselinestretch}{0.8}\hspace*{3em}\begin{minipage}{5em}\footnotesize [\themycommentcounter]: \raggedright ##1\end{minipage}}]{\renewcommand{\baselinestretch}{0.8}\begin{minipage}{5em}\footnotesize [\themycommentcounter]: \raggedright ##1\end{minipage}}}}\definecolor{darkgreen}{RGB}{0,200,100}\definecolor{orange}{RGB}{255,80,0}\newcommand{\Xcomment}[1]{}\newcommand{\skipDTZ}{\phantom{.00}} \newcommand{\enabled}{\ensuremath{\bullet}} \newcommand{\disabled}{\ensuremath{\circ}}       \newcommand{\astar}{A*\xspace} \newcommand{\pot}{\ensuremath{\pi}} \newcommand{\landmarks}{\ensuremath{L}} \newcommand{\landmark}{\ensuremath{l}} \newcommand{\reach}{\ensuremath{r}}  \newcommand{\separator}{\ensuremath{S}} \newcommand{\transitnodes}{\ensuremath{T}} \newcommand{\accessnodes}{\ensuremath{A}}            \newcommand{\reals}{\ensuremath{\mathbb{R}}} \newcommand{\naturals}{\ensuremath{\mathbb{Z}_{\geq 0}}}\newcommand{\graph}{\ensuremath{G}}    \newcommand{\vertices}{\ensuremath{V}}       \newcommand{\edges}{\ensuremath{E}} \newcommand{\arcs}{\ensuremath{A}}     \newcommand{\apath}{\ensuremath{P}}     \newcommand{\searchspace}{\ensuremath{S}}\newcommand{\cost}{\ensuremath{\ell}}       \newcommand{\vertex}{\ensuremath{u}} \newcommand{\vertexa}{\ensuremath{u}} \newcommand{\vertexb}{\ensuremath{v}} \newcommand{\vertexc}{\ensuremath{w}}       \newcommand{\arc}{\ensuremath{a}}      \DeclareMathOperator{\dist}{dist} \newcommand{\partition}{\ensuremath{\mathcal{C}}} \newcommand{\cell}{\ensuremath{C}}                            \newcommand{\sourcevertex}{\ensuremath{s}} \newcommand{\targetvertex}{\ensuremath{t}}\newcommand{\alphabet}{\ensuremath{\Sigma}}    \newcommand{\alanguage}{\ensuremath{L}}   \newcommand{\abs}[1]{\left|#1\right|}   \newcommand{\astate}{\ensuremath{q}}    \newcommand{\bigO}{\ensuremath{\mathcal{O}}}            \newcommand{\maxtransfers}{\ensuremath{K}} \newcommand{\journeys}{\ensuremath{\mathcal{J}}}    \newcommand{\journey}{\ensuremath{J}}     \newcommand{\st}{\ensuremath{p}} \newcommand{\trip}{\ensuremath{t}}  \newcommand{\route}{\ensuremath{r}}        \newcommand{\deptime}{\ensuremath{\tau_{\text{dep}}}}         \newcommand{\timespan}{\ensuremath{\Delta}} \newcommand{\frequency}{\ensuremath{f}}\newcommand{\sourcestop}{\ensuremath{\st_s}} \newcommand{\sourcetime}{\ensuremath{\tau}}  \newcommand{\targetstop}{\ensuremath{\st_t}} \newcommand{\queue}{\ensuremath{Q}}                                \newcommand{\lab}{\ensuremath{L}}                 \newcommand{\mmgraph}{\ensuremath{\boldsymbol{G}}}                                                  \definecolor{kit-gruen}{cmyk/RGB}{1,0,.6,0/0,150,130} \definecolor{kit-gruen70}{cmyk/RGB}{.7,0,.42,0/76,181,167} \definecolor{kit-gruen50}{cmyk/RGB}{.5,0,.3,0/127,202,192} \definecolor{kit-gruen30}{cmyk/RGB}{.3,0,.18,0/178,223,217} \definecolor{kit-gruen15}{cmyk/RGB}{.15,0,.09,0/217,239,236}\definecolor{kit-blau}{cmyk/RGB}{.8,.5,0,0/70,100,170} \definecolor{kit-blau70}{cmyk/RGB}{.56,.35,0,0/125,146,195} \definecolor{kit-blau50}{cmyk/RGB}{.40,.25,0,0/162,177,212} \definecolor{kit-blau30}{cmyk/RGB}{.24,.15,0,0/199,208,229} \definecolor{kit-blau15}{cmyk/RGB}{.12,.075,0,0/227,232,242}\definecolor{kit-schwarz}{cmyk/RGB}{0,0,0,1/0,0,0} \definecolor{kit-schwarz70}{cmyk/RGB}{0,0,0,.7/77,77,77} \definecolor{kit-schwarz50}{cmyk/RGB}{0,0,0,.5/128,128,128} \definecolor{kit-schwarz30}{cmyk/RGB}{0,0,0,.3/179,179,179} \definecolor{kit-schwarz15}{cmyk/RGB}{0,0,0,.15/217,217,217}\definecolor{kit-maigruen}{cmyk/RGB}{.6,0,1,0/92,172,53} \definecolor{kit-gelb}{cmyk/RGB}{0,.05,1,0/225,227,18} \definecolor{kit-orange}{cmyk/RGB}{0,.45,1,0/229,131,35} \definecolor{kit-braun}{cmyk/RGB}{.35,.5,1,0/144,105,43} \definecolor{kit-rot}{cmyk/RGB}{.25,1,1,0/155,23,35} \definecolor{kit-lila}{cmyk/RGB}{.25,1,0,0/152,5,104} \definecolor{kit-cyan-blau}{cmyk/RGB}{.9,.05,0,0/0,144,204} \usepackage{pgfplotstable} \usetikzlibrary{automata,positioning,arrows,calc,shapes,decorations.text,decorations.pathmorphing,external,shapes.callouts,decorations.markings} \usepgfplotslibrary{dateplot}\pgfdeclarelayer{edgelayer} \pgfdeclarelayer{nodelayer} \pgfsetlayers{edgelayer,nodelayer,main} \tikzset{initial text={}} \tikzset{>=stealth'} \tikzset{shorten >=1pt} \tikzset{shorten <=1pt} \tikzstyle{every pin}=[pin distance=.25cm]\tikzstyle{every state}=[minimum size=20pt,draw=kit-schwarz,fill=kit-gruen15] \tikzstyle{loop transition}=[->,kit-schwarz] \tikzstyle{link transition}=[->,kit-blau,thick,every node/.style={fill=white}] \tikzstyle{vertex}=[style={circle,draw=kit-schwarz,fill=kit-gruen15}] \tikzstyle{deleted vertex}=[vertex,style={fill=kit-rot!15}] \tikzstyle{highlighted edge}=[draw,line width=4pt,-,kit-gruen30] \tikzstyle{squiggly edge}=[draw,decorate,decoration={snake,segment length=15pt}] \tikzstyle{squiggly arc}=[draw,->,decorate,decoration={snake,segment length=20pt,post length=3pt}] \tikzstyle{smallvertex}=[style={circle,fill=kit-schwarz70,inner sep=0pt,minimum size=8pt}] \tikzstyle{tinyvertex}=[style={circle,fill=kit-schwarz70,inner sep=0pt,minimum size=5pt}] \tikzstyle{directed edge}=[->]\tikzstyle{plotpoint}=[style={circle,draw,thick,kit-gruen,fill=white,inner sep=0pt,minimum size=5pt}] \tikzstyle{paretopoint}=[plotpoint,style={fill=kit-gruen}]\tikzstyle{arrival vertex}=[style={diamond,draw=kit-schwarz,fill=kit-gelb!50}] \tikzstyle{departure vertex}=[style={rectangle,draw=kit-schwarz,fill=kit-maigruen!50,minimum size=9pt}] \tikzstyle{transfer vertex}=[style={circle,draw=kit-schwarz,fill=kit-lila!50}] \tikzstyle{landmark vertex}=[vertex,fill=kit-gelb!50] \tikzstyle{stay arc}=[->,>=stealth'] \tikzstyle{departure arc}=[->,>=stealth'] \tikzstyle{connection arc}=[->,>=stealth',thick,every node/.style={fill=white}]\tikzstyle{stop vertex}=[style={circle,draw=kit-schwarz,fill=kit-blau!50}] \tikzstyle{airport vertex}=[style={circle,draw=kit-schwarz,fill=kit-blau!50}] \tikzstyle{superstop vertex}=[style={rectangle,rounded corners,draw=kit-schwarz,fill=kit-blau!50,minimum size=12pt}] \tikzstyle{sourcetargetsuperstop vertex}=[style={rectangle,rounded corners,draw=kit-schwarz,fill=kit-braun!15,minimum size=12pt}] \tikzstyle{hubsuperstop vertex}=[style={rectangle,rounded corners,draw=kit-schwarz,thick,fill=kit-blau!50,minimum size=12pt}] \tikzstyle{viasuperstop vertex}=[style={rectangle,rounded corners,draw=kit-schwarz,thick,fill=kit-rot!50,minimum size=12pt}] \tikzstyle{route vertex}=[style={rectangle,draw=kit-schwarz,fill=kit-orange!50,minimum size=9pt}] \tikzstyle{transfer arc}=[->,>=stealth',bend right] \tikzstyle{airport arc}=[->,>=stealth'] \tikzstyle{route arc}=[->,>=stealth',thick,every node/.style={fill=white}] \tikzstyle{squiggly route arc}=[->,thick,draw,decorate,decoration={snake,segment length=20pt,post length=5pt}]\tikzstyle{conflict vertex}=[style={rectangle,rounded corners=3pt,draw=kit-schwarz,fill=kit-gruen15,minimum size=10pt}] \tikzstyle{conflict edge}=[-] \tikzstyle{partition boundary}=[-,very thick,kit-schwarz30] \tikzstyle{cut edge}=[thick,-,kit-rot] \tikzstyle{clique edge}=[-,kit-gruen] \tikzstyle{skeleton vertex}=[tinyvertex,kit-gruen]\tikzstyle{stop}=[style={circle,very thick,draw=white,inner sep=0pt,minimum size=6pt}] \tikzstyle{importantstop}=[style={circle,very thick,draw=white,inner sep=0pt,minimum size=9pt}] \tikzstyle{transferstop}=[style={circle,draw=kit-schwarz,inner sep=0pt,minimum size=6pt,fill=white}] \tikzstyle{labeledstop}=[style={rectangle,rounded corners=3pt,draw=kit-schwarz,fill=white}] \tikzstyle{sourcetargetstop}=[style={rectangle,rounded corners=3pt,draw=kit-schwarz,fill=kit-braun!15}] \tikzstyle{route}=[-,line width=3pt,draw,rounded corners=3pt] \tikzstyle{walking}=[line width=1.5pt,kit-schwarz!90] \tikzstyle{cycling}=[line width=1.5pt,kit-lila] \tikzstyle{squigglyroute}=[-,line width=3pt,draw,decorate,decoration={snake,segment length=20pt}]\pgfmathsetseed{3141592} \newcommand*{\computeNine}{0.43063572427651} \newcommand*{\computeEleven}{1.19509594882729} \newcommand*{\spaTwo}{1.0} \newcommand*{\spaThree}{0.6378694788814} \newcommand*{\chOpteron}{0.37066962847405} \newcommand*{\pfoser}{1.2546346782988}\usetikzlibrary{external}\newcommand{\tabhead}[1]{{\textsc{#1}}}\newcommand{\multitabhead}[2]{\multicolumn{#1}{c}{\tabhead{#2}}}\newcommand{\email}[1]{\texttt{#1}}\title{Route Planning in Transportation Networks\thanks{This is an updated version of the technical report MSR-TR-2014-4, previously published by Microsoft Research. This work was mostly done while the authors Daniel Delling, Andrew Goldberg, and Renato F.~Werneck were at Microsoft Research Silicon Valley.}}\author[1]{Hannah Bast}\author[2]{Daniel Delling}\author[3]{Andrew Goldberg}\author[4]{Matthias~M{\"u}ller-Hannemann}\author[5]{Thomas Pajor}\author[6]{Peter Sanders}\author[7]{Dorothea Wagner}\author[8]{Renato F.~Werneck}\affil[1]{University of Freiburg, Germany, \email{bast@informatik.uni-freiburg.de}}\affil[2]{Sunnyvale, USA, \email{daniel.delling@gmail.com}}\affil[3]{Amazon, USA, \email{andgold@amazon.com}}\affil[4]{Martin-Luther-Universit{\"a}t Halle-Wittenberg, Germany\\\email{muellerh@informatik.uni-halle.de}}\affil[5]{Microsoft Research, USA, \email{tpajor@microsoft.com}}\affil[6]{Karlsruhe Institute of Technology, Germany, \email{sanders@kit.edu}}\affil[7]{Karlsruhe Institute of Technology, Germany, \email{dorothea.wagner@kit.edu}}\affil[8]{Amazon, USA, \email{werneck@amazon.com}}\date{April 17, 2015}
\begin{document} \maketitle \thispagestyle{empty}

\begin{abstract} We survey recent advances in algorithms for route planning in transportation networks. For road networks, we show that one can compute driving directions in milliseconds or less even at continental scale. A variety of techniques provide different trade-offs between preprocessing effort, space requirements, and query time. Some algorithms can answer queries in a fraction of a microsecond, while others can deal efficiently with real-time traffic. Journey planning on public transportation systems, although conceptually similar, is a significantly harder problem due to its inherent time-dependent and multicriteria nature. Although exact algorithms are fast enough for interactive queries on metropolitan transit systems, dealing with continent-sized instances requires simplifications or heavy preprocessing. The multimodal route planning problem, which seeks journeys combining schedule-based transportation~(buses, trains) with unrestricted modes~(walking, driving), is even harder, relying on approximate solutions even for metropolitan inputs. \end{abstract}

\clearpage

\section{Introduction} \label{sec:introduction}

This survey is an introduction to the state of the art in the area of practical algorithms for routing in transportation networks. Although a thorough survey by Delling et al.~\cite{dssw-erpa-09} has appeared fairly recently, it has become outdated due to significant developments in the last half-decade. For example, for continent-sized road networks, newly-developed algorithms can answer queries in a few hundred nanoseconds; others can incorporate current traffic information in under a second on a commodity server; and many new applications can now be dealt with efficiently. While Delling et al.~focused mostly on road networks, this survey has a broader scope, also including schedule-based public transportation networks as well as multimodal scenarios~(combining schedule-based and unrestricted modes).

Section~\ref{sec:p2p} considers shortest path algorithms for static networks; although it focuses on methods that work well on road networks, they can be applied to arbitrary graphs. Section~\ref{sec:road} then considers the relative performance of these algorithms on real road networks, as well as how they can deal with other transportation applications. Despite recent advances in routing in road networks, there is still no ``best'' solution for the problems we study, since solution methods must be evaluated according to different measures. They provide different trade-offs in terms of query times, preprocessing effort, space usage, and robustness to input changes, among other factors. While solution quality was an important factor when comparing early algorithms, it is no longer an issue: as we shall see, all current state-of-the-art algorithms find provably exact solutions. In this survey, we focus on algorithms that are not clearly dominated by others. We also discuss approaches that were close to the dominance frontier when they were first developed, and influenced subsequent algorithms.

Section~\ref{sec:public} considers algorithms for journey planning on schedule-based public transportation systems~(consisting of buses, trains, and trams, for example), which is quite different from routing in road networks. Public transit systems have a time-dependent component, so we must consider multiple criteria for meaningful results, and known preprocessing techniques are not nearly as effective. Approximations are thus sometimes still necessary to achieve acceptable performance. Advances in this area have been no less remarkable, however: in a few milliseconds, it is now possible to find good journeys within public transportation systems at a very large scale.

Section~\ref{sec:mm} then considers a true \emph{multimodal} scenario, which combines schedule-based means of transportation with less restricted ones, such as walking and cycling. This problem is significantly harder than its individual components, but reasonable solutions can still be found.

A distinguishing feature of the methods we discuss in this survey is that they quickly made real-life impact, addressing problems that need to be solved by interactive systems at a large scale. This demand facilitated technology transfer from research prototypes to practice. As our concluding remarks~(Section~\ref{sec:conclusion}) will explain, several algorithms we discuss have found their way into mainstream production systems serving millions of users on a daily basis.

This survey considers research published until January 2015. We refer to the final~(journal) version of a result, citing conference publications only if a journal version is not yet available. The reader should keep in mind that the journal publications we cite often report on work that first appeared~(at a conference) much earlier.

\section{Shortest Paths Algorithms} \label{sec:p2p}

Let~$\graph = (\vertices,\arcs)$ be a~(directed) graph with a set~$\vertices$ of vertices and a set~$\arcs$ of arcs. Each arc~$(\vertexa,\vertexb) \in A$ has an associated nonnegative \emph{length}~$\cost(\vertexa,\vertexb)$. The length of a path is the sum of its arc lengths. In the \emph{point-to-point shortest path problem}, one is given as input the graph~$\graph$, a source~$\sourcevertex \in \vertices$, and a target~$\targetvertex \in \vertices$, and must compute the length of the shortest path from~$\sourcevertex$ to~$\targetvertex$ in~$\graph$. This is also denoted as~$\dist(\sourcevertex,\targetvertex)$, the distance between~$\sourcevertex$ and~$\targetvertex$. The \emph{one-to-all} problem is to compute the distances from a given vertex~$\sourcevertex$ to all vertices of the graph. The \emph{all-to-one} problem is to find the distances from all vertices to~$\sourcevertex$. The \emph{many-to-many} problem is as follows: given a set~$S$ of sources and a set~$T$ of targets, find the distances~$\dist(\sourcevertex,\targetvertex)$ for all~$\sourcevertex \in S$,~$\targetvertex \in T$. For~$S = T = \vertices$ we have the \emph{all pairs shortest path} problem.

In addition to the distances, many applications need to find the corresponding shortest paths. An \emph{out-shortest path tree} is a compact representation of one-to-all shortest paths from the root~$r$. (Likewise, the in-shortest path tree represents the all-to-one paths.) For each vertex~$\vertex \in \vertices$, the path from~$r$ to~$\vertex$ in the tree is the shortest path.

In this section, we focus on the basic point-to-point shortest path problem under the basic \emph{server model}. We assume that all data fits in RAM. However, locality matters, and algorithms with fewer cache misses run faster. For some algorithms, we consider multi-core and machine-tailored implementations. In our model, preprocessing may be performed on a more powerful machine than queries~(e.\,g., a machine with more memory). While preprocessing may take a long time~(e.\,g., hours), queries need to be fast enough for interactive applications.

In this section, we first discuss basic techniques, then those using preprocessing. Since all methods discussed could in principle be applied to arbitrary graphs, we keep the description as general as possible. For intuition, however, it pays to keep road networks in mind, considering that they were the motivating application for most approaches we consider. We will explicitly consider road networks, including precise performance numbers, in Section~\ref{sec:road}.

\subsection{Basic Techniques}

\begin{figure}[t] \begin{minipage}{\textwidth} \centering \raisebox{-.5\height}{\begin{tikzpicture}[scale=0.66] \def\pts{} \coordinate (TRGT) at (0:2.15); \foreach \x in {1,...,9} { \xdef\pts{\pts (\x*360/10-5+10*rnd:2+.3*rnd)}; } \path[fill=kit-gruen,fill opacity=.15] plot [smooth cycle, tension=0.8] coordinates { (TRGT) \pts }; \node[smallvertex] (T) at (TRGT) [label=right:$\targetvertex$] {}; \node[smallvertex] (S) at (0,0) [label=above:$\sourcevertex$] {}; \path[squiggly arc] (S) -- (T); \end{tikzpicture} } \hfill \raisebox{-.5\height}{\begin{tikzpicture}[scale=.85] \def\spts{} \def\tpts{} \foreach \x in {1,...,9} { \xdef\spts{\spts (\x*360/10+10+10*rnd:1+.3*rnd)}; } \foreach \x in {1,...,9} { \xdef\tpts{\tpts (\x*360/10+10+10*rnd:1+.3*rnd)}; } \path[fill=kit-gruen,fill opacity=0.15] plot [smooth cycle, tension=0.8] coordinates { (15:1.15) \spts }; \begin{scope}[xshift=2.2cm,xscale=-1] \path[fill=kit-rot,fill opacity=0.15] plot [smooth cycle, tension=0.8] coordinates { (15:1.15) \tpts }; \node[smallvertex] (T) at (0,0) [label=above:$\targetvertex$] {}; \end{scope} \node[smallvertex] (S) at (0,0) [label=above:$\sourcevertex$] {}; \node[smallvertex] (M) at (15:1.15) [pin=90:$x$] {}; \path[squiggly arc] (S) -- (M); \path[squiggly arc] (M) -- (T); \end{tikzpicture} } \hfill \raisebox{-.5\height}{\begin{tikzpicture}[scale=.66] \def\eggheight{2cm} \path[rotate=270, fill=kit-gruen, fill opacity=.15] plot[domain=-pi:pi,samples=100] ({.6*\eggheight *cos(\x/4 r)*sin(\x r)},{-\eggheight*(cos(\x r))}) -- cycle; \node[smallvertex] (S) at ({0.5cm-\eggheight},0) [label=above:$\sourcevertex$] {}; \node[smallvertex] (T) at (\eggheight,0) [label=right:$\targetvertex$] {}; \path[squiggly arc] (S) -- (T); \end{tikzpicture} } \end{minipage} \caption{Schematic search spaces of Dijkstra's algorithm~(left), bidirectional search~(middle), and the \astar algorithm~(right).} \label{fig:literature:dijbisearchastar} \end{figure}
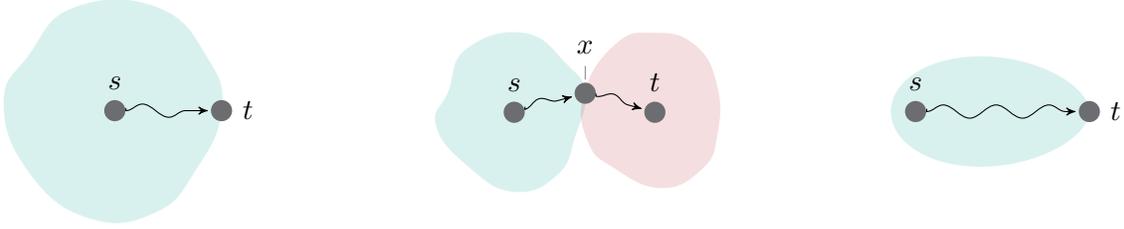

The standard solution to the one-to-all shortest path problem is Dijkstra's algorithm~\cite{d-ntpcg-59}. It maintains a priority queue~$\queue$ of vertices ordered by~(tentative) distances from~$\sourcevertex$. The algorithm initializes all distances to infinity, except~$\dist(\sourcevertex,\sourcevertex) = 0$, and adds~$\sourcevertex$ to~$\queue$. In each iteration, it extracts a vertex~$\vertexa$ with minimum distance from~$\queue$ and \emph{scans} it, i.\,e., looks at all arcs~$\arc = (\vertexa,\vertexb) \in \arcs$ incident to~$\vertexa$. For each such arc, it determines the distance to~$\vertexb$ via arc~$\arc$ by computing~$\dist(\sourcevertex,\vertexa) + \cost(\arc)$. If this value improves~$\dist(\sourcevertex,\vertexb)$, the algorithm performs an arc \emph{relaxation}: it updates~$\dist(\sourcevertex,\vertexb)$ and adds vertex~$\vertexb$ with key~$\dist(\sourcevertex,\vertexb)$ to the priority queue~$\queue$. Dijkstra's algorithm has the \emph{label-setting} property: once a vertex~$\vertex \in \vertices$ is scanned~(settled), its distance value~$\dist(\sourcevertex,\vertex)$ is correct. Therefore, for point-to-point queries, the algorithm may stop as soon as it scans the target~$\targetvertex$. We refer to the set of vertices~$\searchspace \subseteq \vertices$ scanned by the algorithm as its \emph{search space}. See \figurename~\ref{fig:literature:dijbisearchastar} for an illustration.

The running time of Dijkstra's algorithm depends on the priority queue used. The running time is~$\bigO((\abs{\vertices} + \abs{\arcs})\log\abs{\vertices})$ with binary heaps~\cite{w-ah-64}, improving to~$\bigO(\abs{\arcs} + \abs{\vertices}\log\abs{\vertices})$ with Fibonacci heaps~\cite{ft-fhtui-87}. For arbitrary~(non-integral) costs, generalized versions of binary heaps~(such as 4-heaps or 8-heaps) tend to work best in practice~\cite{cgr-s-96}. If all arc costs are integers in the range~$[0,C]$, multi-level buckets~\cite{df-srmrp-79} yield a running time of~$\bigO(\abs{\arcs}+\abs{\vertices}\sqrt{\log C})$~\cite{amot-faspp-90,cgs-bhlmp-97} and work well in practice. For the average case, one can get an~$\bigO(\abs{\vertices} + \abs{\arcs})$~(linear) time bound~\cite{m-ssspa-01,g-apspa-08}. Thorup~\cite{Thorup03} has improved the theoretical worst-case bound of Dijkstra's algorithm to~$O(\abs{\arcs} + \abs{\vertices}\log\log\min\{\abs{\vertices},C\})$, but the required data structure is rather involved and unlikely to be faster in practice.

In practice, one can reduce the search space using \emph{bidirectional search}~\cite{d-lpe-62}, which simultaneously runs a forward search from~$\sourcevertex$ and a backward search from~$\targetvertex$. The algorithm may stop as soon as the intersection of their search spaces provably contains a vertex~$x$ on the shortest path from~$\sourcevertex$ to~$\targetvertex$. For road networks, bidirectional search visits roughly half as many vertices as the unidirectional approach.

An alternative method for computing shortest paths is the Bellman-Ford algorithm~\cite{f-nft-56,b-orp-58,moo-59}. It uses no priority queue. Instead, it works in rounds, each scanning all vertices whose distance labels have improved. A simple FIFO queue can be used to keep track of vertices to scan next. It is a \emph{label-correcting} algorithm, since each vertex may be scanned multiple times. Although it runs in~$\bigO(\abs{\vertices}\abs{\arcs})$ time in the worst case, it is often much faster, making it competitive with Dijkstra's algorithm in some scenarios. In addition, it works on graphs with negative edge weights.

Finally, the Floyd-Warshall algorithm~\cite{f-asp-62} computes distances between \emph{all} pairs of vertices in~$\Theta(\abs{\vertices}^3)$ time. For sufficiently dense graphs, this is faster than~$\abs{\vertices}$ calls to Dijkstra's algorithm.

\subsection{Goal-Directed Techniques}

Dijkstra's algorithm scans all vertices with distances smaller than~$\dist(\sourcevertex,\targetvertex)$. Goal-directed techniques, in contrast, aim to ``guide'' the search toward the target by avoiding the scans of vertices that are not in the direction of~$\targetvertex$. They either exploit the~(geometric) embedding of the network or properties of the graph itself, such as the structure of shortest path trees toward~(compact) regions of the graph.

\paragraph{\astar Search.}

A classic goal-directed shortest path algorithm is \astar~search~\cite{hnr-afbhd-68}. It uses a potential function~$\pot \colon \vertices \to \reals$ on the vertices, which is a \emph{lower bound} on the distance~$\dist(\vertex,\targetvertex)$ from~$\vertex$ to~$\targetvertex$. It then runs a modified version of Dijkstra's algorithm in which the priority of a vertex~$\vertexa$ is set to~$\dist(\sourcevertex,\vertexa) + \pot(\vertexa)$. This causes vertices that are closer to the target~$\targetvertex$ to be scanned earlier during the algorithm. See~\figurename~\ref{fig:literature:dijbisearchastar}. In particular, if~$\pot$ were an \emph{exact} lower bound~($\pot(\vertex) = \dist(\vertex,\targetvertex)$), \emph{only} vertices along shortest~$\sourcevertex$--$\targetvertex$ paths would be scanned. More vertices may be visited in general but, as long as the potential function is \emph{feasible}~(i.\,e., if~$\cost(\vertexb,\vertexc) - \pot(\vertexb) + \pot(\vertexc) \geq 0$ for~$(\vertexb,\vertexc) \in \edges$), an~$\sourcevertex$--$\targetvertex$ query can stop with the correct answer as soon as it is about to scan the target vertex~$\targetvertex$.

The algorithm can be made bidirectional, but some care is required to ensure correctness. A standard approach is to ensure that the forward and backward potential functions are consistent. In particular, one can combine two arbitrary feasible functions~$\pot_f$ and~$\pot_r$ into consistent potentials by using ~$(\pot_f - \pot_r)/2$ for the forward search and~$(\pot_r - \pot_f)/2$ for the backward search~\cite{ihinshtm-aai-94}. Another approach, which leads to similar results in practice, is to change the stopping criterion instead of making the two functions consistent~\cite{p-bdhsp-69,kk-bhsr-97,gh-cspas-05,rt-basaa-12}.

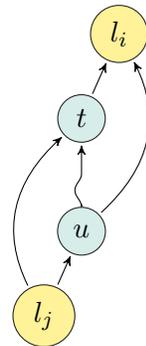
\begin{wrapfigure}{O}{0pt} \begin{tikzpicture}[yscale=.75] \node[vertex] (V) at (0,0) {$\vertexa$}; \node[vertex] (T) at (0,2) {$\targetvertex$}; \node[landmark vertex] (L1) at (.5,3.5) {$\landmark_i$}; \node[landmark vertex] (L2) at (-.5,-1.5) {$\landmark_j$}; \path[squiggly arc] (V) -- (T); \path[directed edge] (T) edge (L1); \path[directed edge] (V) edge [bend right] (L1); \path[directed edge] (L2) edge [bend left] (T); \path[directed edge] (L2) edge (V); \end{tikzpicture} \caption[Illustration of the triangle inequalities for ALT]{Triangle inequalities for ALT.} \label{fig:literature:alt} \end{wrapfigure}

In road networks with travel time metric, one can use the geographical distance~\cite{p-bs-71,sv-speg-86} between~$\vertex$ and~$\targetvertex$ divided by the maximum travel speed~(that occurs in the network) as the potential function. Unfortunately, the corresponding bounds are poor, and the performance gain is small or non-existent~\cite{gh-cspas-05}. In practice, the algorithm can be accelerated using more aggressive bounds~(for example, a smaller denominator), but correctness is no longer guaranteed. In practice, even when minimizing travel distances in road networks, \astar with geographical distance bound performs poorly compared to other modern methods.

One can obtain much better lower bounds~(and preserve correctness) with the \emph{ALT}~(\emph{\astar, landmarks, and triangle inequality}) algorithm~\cite{gh-cspas-05}. During a preprocessing phase, it picks a small set~$\landmarks \subseteq \vertices$ of \emph{landmarks} and stores the distances between them and all vertices in the graph. During an~$\sourcevertex$--$\targetvertex$ query, it uses triangle inequalities involving the landmarks to compute a valid lower bound on~$\dist(\vertexa,\targetvertex)$ for any vertex~$\vertexa$. More precisely, for any landmark~$\landmark_i$, both~$\dist(\vertexa,\targetvertex) \geq \dist(\vertexa,\landmark_i) - \dist(\targetvertex,\landmark_i)$ and~$\dist(\vertexa,\targetvertex) \geq \dist(\landmark_i,\targetvertex) - \dist(\landmark_i,\vertexa)$ hold. If several landmarks are available, one can take the maximum overall bound. See~\figurename~\ref{fig:literature:alt} for an illustration. The corresponding potential function is feasible~\cite{gh-cspas-05}.

The quality of the lower bounds~(and thus query performance) depends on which vertices are chosen as landmarks during preprocessing. In road networks, picking well-spaced landmarks close to the boundary of the graph leads to the best results, with acceptable query times on average~\cite{gw-cppsp-05,ep-olbrp-13}. For a small~(but noticeable) fraction of the queries, however, speedups relative to bidirectional Dijkstra are minor.

\paragraph{Geometric Containers.}

Another goal-directed method is \emph{Geometric Containers}. It precomputes, for each arc~$\arc = (\vertexa,\vertexb) \in \arcs$, an arc label~$\lab(\arc)$ that encodes the set~$\vertices_\arc$ of vertices to which a shortest path from~$\vertexa$ begins with the arc~$\arc$. Instead of storing~$\vertices_\arc$ explicitly,~$\lab(\arc)$ approximates this set by using geometric information~(i.\,e., the coordinates) of the vertices in~$\vertices_\arc$. During a query, if the target vertex~$\targetvertex$ is not in~$\lab(\arc)$, the search can safely be pruned at~$\arc$. Schulz et al.~\cite{sww-daola-00} approximate the set~$\vertices_\arc$ by an angular sector~(centered at~$\vertexa$) that covers all vertices in~$\vertices_\arc$. Wagner et al.~\cite{wwz-gcesp-05} consider other geometric containers, such as ellipses and the convex hull, and conclude that bounding boxes perform consistently well. For graphs with no geometric information, one can use graph layout algorithms and then create the containers~\cite{bsww-tpsmm-01,ww-dgsus-05}. A disadvantage of Geometric Containers is that its preprocessing essentially requires an all-pairs shortest path computation, which is costly.

\paragraph{Arc Flags.} The \emph{Arc Flags} approach~\cite{l-aeept-09,hkms-fppsp-09} is somewhat similar to Geometric Containers, but does not use geometry. During preprocessing, it partitions the graph into~$K$ cells that are roughly \emph{balanced}~(have similar number of vertices) and have a small number of boundary vertices. Each arc maintains a vector of~$K$ bits~(arc flags), where the~$i$-th bit is set if the arc lies on a shortest path to some vertex of cell~$i$. The search algorithm then prunes arcs which do not have the bit set for the cell containing~$\targetvertex$. For better query performance, arc flags can be extended to nested multilevel partitions~\cite{mssww-pgsda-06}. Whenever the search reaches the cell that contains~$\targetvertex$, it starts evaluating arc flags with respect to the~(finer) cells of the level below. This approach works best in combination with bidirectional search~\cite{hkms-fppsp-09}.

The arc flags for a cell~$i$ are computed by growing a backward shortest path tree from each boundary vertex~(of cell~$i$), setting the~$i$-th flag for all arcs of the tree. Alternatively, one can compute arc flags by running a label-correcting algorithm from all boundary vertices simultaneously~\cite{hkms-fppsp-09}. To reduce preprocessing space, one can use a compression scheme that flips some flags from zero to one~\cite{bdgw-sesr-10}, which preserves correctness. As Section~\ref{sec:road} will show, Arc Flags currently have the fastest query times among purely goal-directed methods for road networks. Although high preprocessing times~(of several hours) have long been a drawback of Arc Flags, the recent PHAST algorithm~(cf.~Section~\ref{ssec:extensions}) can make this method more competitive with other techniques~\cite{dgnw-phast-13}.

\paragraph{Precomputed Cluster Distances.} Another goal-directed technique is \emph{Precomputed Cluster Distances}~(PCD)~\cite{msm-gdspq-09}. Like Arc Flags, it is based on a~(preferably balanced) partition~$\partition = (\cell_1, \ldots, \cell_K)$ with~$K$ cells~(or clusters). The preprocessing algorithm computes the shortest path distances between all pairs of cells.

The query algorithm is a pruned version of Dijkstra's algorithm. For any vertex~$\vertexa$ visited by the search, a valid lower bound on its distance to the target is~$\dist(\sourcevertex,\vertexa) + \dist(\cell(\vertexa),\cell(\targetvertex)) + \dist(\vertexb,\targetvertex)$, where~$\cell(\vertex)$ is the cell containing~$\vertex$ and~$\vertexb$ is the boundary vertex of~$\cell(\targetvertex)$ that is closest to~$\targetvertex$. If this bound exceeds the best current upper bound on~$\dist(\sourcevertex,\targetvertex)$, the search is pruned. For road networks, PCD has similar query times to ALT, but requires less space.

\paragraph{Compressed Path Databases.} The Compressed Path Databases~(CPD)~\cite{b-uoprs-11,bh-ppcap-13} method implicitly stores all-pairs shortest path information so that shortest paths can be quickly retrieved during queries. Each vertex~$\vertexa \in \vertices$ maintains a label~$\lab(\vertex)$ that stores the \emph{first move}~(the arc incident to~$\vertexa$) of the shortest path toward \emph{every} other vertex~$\vertexb$ of the graph. A query from~$\sourcevertex$ simply scans~$\lab(\vertex)$ for~$\targetvertex$, finding the first arc~$(\sourcevertex,\vertex)$ of the shortest path~(to~$\targetvertex$); it then recurses on~$\vertex$ until it reaches~$\targetvertex$. Explicitly storing the first arc of every shortest path~(in~$\Theta(\abs{\vertices}^2)$ space) would be prohibitive. Instead, Botea and Harabor~\cite{bh-ppcap-13} propose a lossless data compression scheme that groups vertices that share the same first move~(out of~$\vertex$) into nonoverlapping geometric rectangles, which are then stored with~$\vertex$. Further optimizations include storing the most frequent first move as a default and using more sophisticated compression techniques. This leads to fast queries, but space consumption can be quite large; the method is thus dominated by other approaches. CPD can be seen as an evolution of the \emph{Spatially Induced Linkage Cognizance}~(SILC) algorithm~\cite{sas-eqpsn-05}, and both can be seen as stronger versions of Geometric Containers.

\subsection{Separator-Based Techniques} \label{sec:literature:road:separator}

\begin{figure}[t] \centering \begin{tikzpicture}[scale=.66] \def\sly{.5} \def\slx{-1} \def\shy{2.8cm} \begin{scope}[yslant=\sly,xslant=\slx,every node/.append style={ yslant=\sly,xslant=\slx}] \draw[kit-schwarz!50, fill=kit-orange!2.5, fill opacity=.75] (0,0) rectangle (5,2.5); \draw[step=5mm, kit-orange!50] (0,0) grid (5,2.5); \begin{scope}[xshift=.7cm,yshift=1.3cm] \path[fill=kit-gruen!15] (0,0) circle (.5cm); \node[smallvertex] (111) at (100:.3) {}; \node[smallvertex] (112) at (280:.2) {}; \end{scope} \begin{scope}[xshift=2.2cm, yshift=0.8cm] \path[fill=kit-gruen!15] (0,0) circle (.68cm); \node[smallvertex] (121) at (70:.3) {}; \node[smallvertex] (122) at (142:.5) {}; \node[smallvertex] (123) at (310:.4) {}; \end{scope} \begin{scope}[xshift=3.6cm, yshift=1.8cm] \path[fill=kit-gruen!15] (0,0) circle (.56cm); \node[smallvertex] (131) at (45:.3) {}; \node[smallvertex] (132) at (270:.35) {}; \end{scope} \begin{scope}[xshift=4.3cm, yshift=0.65cm] \path[fill=kit-gruen!15] (0,0) circle (.4cm); \node[smallvertex] (141) at (0,0) {}; \end{scope} \end{scope} \begin{scope}[yshift=\shy,yslant=\sly,xslant=\slx,every node/.append style={ yslant=\sly,xslant=\slx}] \coordinate (21) at (1.35,1.25); \coordinate (22) at (2.85,1.5) {}; \coordinate (23) at (4,1.25) {}; \end{scope} \begin{scope}[every edge/.style={draw,-,kit-schwarz!90}] \path (21) edge (111) edge (112) edge (121) edge (122) edge (123); \path (22) edge (121) edge (122) edge (123) edge (131) edge (132); \path (23) edge (131) edge (132) edge (141); \end{scope} \begin{scope}[yshift=\shy,yslant=\sly,xslant=\slx,every node/.append style={yslant=\sly,xslant=\slx}] \draw[kit-schwarz!50, fill=kit-maigruen!2.5, fill opacity=.75] (0,0) rectangle (5,2.5); \draw[step=5mm, kit-maigruen!50] (0,0) grid (5,2.5); \node[smallvertex] (21) at (21) {}; \node[smallvertex] (22) at (22) {}; \node[smallvertex] (23) at (23) {}; \end{scope} \begin{scope}[every edge/.style={draw,-,thick,kit-schwarz!90}] \path (21) edge (22) edge [bend right,dotted] (23) (22) edge (23); \end{scope} \end{tikzpicture} \hfil \begin{tikzpicture}[scale=.66] \newcommand{\drawclique}[5]{ \begin{scope}[xshift=#4,yshift=#5] \def\pts{} \foreach \x in {1,...,#2} { \pgfmathsetmacro{\ang}{\x*360/#2-5+10*rnd} \pgfmathsetmacro{\radi}{#3+.3*rnd} \node[smallvertex] (#1\x) at (\ang:\radi-.25) {}; \xdef\pts{\pts (\ang:\radi)}; \foreach \y in {1,...,\x} { \path (#1\x) edge (#1\y); } } \path[fill=kit-gruen, fill opacity=.15] plot [smooth cycle, tension=0.8] coordinates { \pts }; \end{scope} } \drawclique{A}{6}{1}{0}{0} \drawclique{B}{8}{1.3}{3cm}{0} \drawclique{C}{5}{1}{1.3cm}{-2.6cm} \drawclique{D}{7}{1.2}{4.3cm}{-3.1cm} \path[ultra thick,kit-rot] (A4) edge (C2) (A6) edge (B4) (A5) edge (C2) (C1) edge (B5) (C5) edge (B6) (D3) edge (C5) (D2) edge (B6) (D2) edge (B7) (C4) edge (D4); \end{tikzpicture} \caption{Left: Multilevel overlay graph with two levels. The dots depict separator vertices in the lower and upper level. Right: Overlay graph constructed from arc separators. Each cell contains a full clique between its boundary vertices, and cut arcs are thicker.} \label{fig:literature:separators} \end{figure}
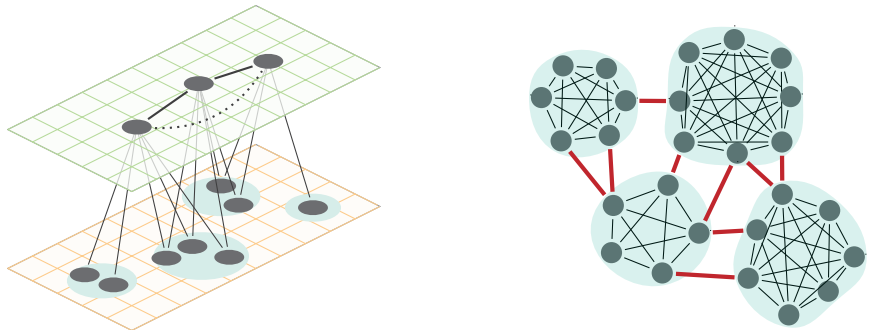

Planar graphs have small~(and efficiently-computable) separators~\cite{lt-astpg-79}. Although road networks are not planar~(think of tunnels or overpasses), they have been observed to have small separators as well~\cite{eg-s-08,dgrw-gpnc-11,ss-degp-12}. This fact is exploited by the methods in this section.

\paragraph{Vertex Separators.}

We first consider algorithms based on~\emph{vertex separators}. A vertex separator is a~(preferably small) subset~$\separator \subset \vertices$ of the vertices whose removal decomposes the graph~$\graph$ into several~(preferably balanced) cells~(components). This separator can be used to compute an~\emph{overlay graph}~$\graph'$ over~$\separator$. Shortcut arcs~\cite{v-ispat-78} are added to the overlay such that distances between \emph{any} pair of vertices from~$\separator$ are preserved, i.\,e., they are equivalent to the distance in~$\graph$. The much smaller overlay graph can then be used to accelerate~(parts of) the query algorithm.

Schulz et al.~\cite{sww-daola-00} use an overlay graph over a carefully chosen subset~$\separator$~(not necessarily a separator) of ``important'' vertices. For each pair of vertices~$\vertexa,\vertexb \in \separator$, an arc~$(\vertexa,\vertexb)$ is added to the overlay if the shortest path from~$\vertexa$ to~$\vertexb$ in~$\graph$ does not contain any other vertex~$\vertexc$ from~$\separator$. This approach can be further extended~\cite{swz-umlgt-02,hsw-emlog-08} to multilevel hierarchies. In addition to arcs between separator vertices of the same level, the overlay contains, for each cell on level~$i$, arcs between the confining level~$i$ separator vertices and the interior level $(i-1)$ separator vertices. See~\figurename~\ref{fig:literature:separators} for an illustration.

Other variants of this approach offer different trade-offs by adding many more shortcuts to the graph during preprocessing, sometimes across different levels~\cite{jhr-hepvp-98,gsvg-psd-98}. In particular \emph{High-Performance Multilevel Routing}~(HPML)~\cite{dhmsw-hpmlr-09} substantially reduces query times but significantly increases the total space usage and preprocessing time. A similar approach, based on path separators for planar graphs, was proposed by Thorup~\cite{t-corad-04} and implemented by Muller and Zachariasen~\cite{mz-fcoad-07}. It works reasonably well to find approximate shortest paths on undirected, planarized versions of road networks.

\paragraph{Arc Separators.}

The second class of algorithms we consider uses \emph{arc separators} to build the overlay graphs. In a first step, one computes a partition~$\partition = (\cell_1, \ldots, \cell_k)$ of the vertices into balanced cells while attempting to minimize the number of cut arcs~(which connect boundary vertices of different cells). Shortcuts are then added to preserve the distances between the boundary vertices within each cell.

An early version of this approach is the \emph{Hierarchical MulTi}~(HiTi) method~\cite{jp-aepcm-02}. It builds an overlay graph containing all boundary vertices and all cut arcs. In addition, for each pair~$\vertexa,\vertexb$ of boundary vertices in~$\cell_i$, HiTi adds to the overlay a shortcut~$(\vertexa,\vertexb)$ representing the shortest path from~$\vertexa$ to~$\vertexb$ in~$\graph$ restricted to~$\cell_i$. The query algorithm then~(implicitly) runs Dijkstra's algorithm on the subgraph induced by the cells containing~$\sourcevertex$ and~$\targetvertex$ plus the overlay. This approach can be extended to use nested multilevel partitions. HiTi has only been tested on grid graphs~\cite{jp-aepcm-02}, leading to modest speedups. See also Figure~\ref{fig:literature:separators}.

The recent \emph{Customizable Route Planning}~(CRP)~\cite{dgpw-crp-11,dgpw-crprn-13} algorithm uses a similar approach, but is specifically engineered to meet the requirements of real-world systems operating on road networks. In particular, it can handle turn costs and is optimized for fast updates of the cost function~(metric). Moreover, it uses PUNCH~\cite{dgrw-gpnc-11}, a graph partitioning algorithm tailored to road networks. Finally, CRP splits preprocessing in two phases: metric-independent preprocessing and customization. The first phase computes, besides the multilevel partition, the topology of the overlays, which are represented as matrices in contiguous memory for efficiency. Note that the partition does not depend on the cost function. The second phase~(which takes the cost function as input) computes the costs of the clique arcs by processing the cells in bottom-up fashion and in parallel. To process a cell, it suffices to run Dijkstra's algorithm from each boundary vertex, but the second phase is even faster using the Bellman-Ford algorithm paired with~(metric-independent) contraction~\cite{dw-fcrn-13}~(cf.~Section~\ref{sec:literature:road:hierarchical}), at the cost of increased space usage. Further acceleration is possible using GPUs~\cite{dkw-cddgp-14}. Queries are bidirectional searches in the overlay graph, as in HiTi.

\subsection{Hierarchical Techniques} \label{sec:literature:road:hierarchical}

Hierarchical methods aim to exploit the inherent hierarchy of road networks. Sufficiently long shortest paths eventually converge to a small arterial network of important roads, such as highways. Intuitively, once the query algorithm is far from the source and target, it suffices to only scan vertices of this subnetwork. In fact, using input-defined road categories in this way is a popular heuristic~\cite{epv-edmss-11,hm-sbrp-11}, though there is no guarantee that it will find exact shortest paths. Fu et al.~\cite{fsr-hspatasa-06} give an overview of early approaches using this technique. Since the algorithms we discuss must find exact shortest paths, their correctness must not rely on unverifiable properties such as input classifications. Instead, they use the preprocessing phase to compute the importance of vertices or arcs according to the actual shortest path structure.

\paragraph{Contraction Hierarchies.} An important approach to exploiting the hierarchy is to use \emph{shortcuts}. Intuitively, one would like to augment~$\graph$ with shortcuts that could be used by long-distance queries to skip over ``unimportant'' vertices.

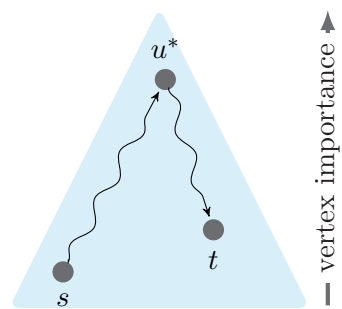
\begin{wrapfigure}{O}{0pt} \begin{tikzpicture}[scale=0.8] \path[fill=kit-cyan-blau!15, rounded corners] (0,0) -- (5,0) -- (2.5,5) -- cycle; \node[smallvertex] (S) at (0.9,0.6) [label=below:$\sourcevertex$] {}; \node[smallvertex] (T) at (3.4,1.3) [label=below:$\targetvertex$] {}; \node[smallvertex] (M) at (2.6,3.8) [label=above:$u^*$] {}; \path[squiggly arc] (S) -- (M); \path[squiggly arc] (M) -- (T); \path[->,>=latex,ultra thick,kit-schwarz!70] (5.3,0) edge node [fill=white,text=kit-schwarz,sloped,anchor=center,auto=false] {vertex importance} (5.3,5); \end{tikzpicture} \caption[Illustration of a Contraction Hierarchies query] {Illustrating a Contraction Hierarchies query.} \label{fig:literature:ch} \end{wrapfigure}

The \emph{Contraction Hierarchies}~(CH) algorithm, proposed by Geisberger et al.~\cite{gssv-erlrn-12}, implements this idea by repeatedly executing a \emph{vertex contraction} operation. To contract a vertex~$\vertexb$, it is~(temporarily) removed from~$\graph$, and a shortcut is created between each pair~$\vertexa,\vertexc$ of neighboring vertices if the shortest path from~$\vertexa$ to~$\vertexc$ is unique and contains~$\vertexb$. During preprocessing, CH~(heuristically) orders the vertices by ``importance'' and contracts them from least to most important.

The query stage runs a bidirectional search from~$\sourcevertex$ and~$\targetvertex$ on~$\graph$ augmented by the shortcuts computed during preprocessing, but only visits arcs leading to vertices of higher ranks~(importance). See \figurename~\ref{fig:literature:ch} for an illustration. Let~$d_\sourcevertex(\vertex)$ and~$d_\targetvertex(\vertex)$ be the corresponding distance labels obtained by these \emph{upward} searches~(set to~$\infty$ for vertices that are not visited). It is easy to show that~$d_\sourcevertex(\vertex) \geq \dist(\sourcevertex,\vertex)$ and~$d_\targetvertex(\vertex) \geq \dist(\vertex,\targetvertex)$; equality is not guaranteed due to pruning. Nevertheless, Geisberger et al.~\cite{gssv-erlrn-12} prove that the highest-ranked vertex~$\vertex^*$ on the original~$\sourcevertex$--$\targetvertex$ path will be visited by both searches, and that both its labels will be exact, i.\,e.,~$d_\sourcevertex(\vertex^*) = \dist(\sourcevertex,\vertex^*)$ and~$d_\targetvertex(\vertex^*) = \dist(\vertex^*,\targetvertex)$. Therefore, among all vertices~$\vertex$ visited by both searches, the one minimizing~$d_\sourcevertex(\vertex) + d_\targetvertex(\vertex)$ represents the shortest path. Note that, since~$\vertex^*$ is not necessarily the first vertex that is scanned by both searches, they cannot stop as soon as they meet.

Query times depend on the vertex order. During preprocessing, the vertex order is usually determined online and bottom-up. The overall~(heuristic) goal is to minimize the number of edges added during preprocessing. One typically selects the vertex to be contracted next by considering a combination of several factors, including the net number of shortcuts added and the number of nearby vertices already contracted~\cite{gssv-erlrn-12,klsv-dtdch-10}. Better vertex orders can be obtained by combining the bottom-up algorithm with~(more expensive) top-down offline algorithms that explicitly classify vertices hitting many shortest paths as more important~\cite{adgw-hhlsp-12,dgpw-rdqmn-14}. Since road networks have very small separators~\cite{dgrw-gpnc-11}, one can use nested dissection to obtain reasonably good orders that work for any length function~\cite{dw-fcrn-13,dsw-cch-sea-14}. Approximate CH has been considered as a way to accommodate networks with less inherent hierarchy~\cite{gs-hchag-10}.

CH is actually a successor of Highway Hierarchies~\cite{ss-ehh-12} and Highway Node Routing~\cite{ss-dhnr-07}, which are based on similar ideas. CH is not only faster, but also conceptually simpler. This simplicity has made it quite versatile, serving as a building block not only for other point-to-point algorithms~\cite{als-tnrr-13,adgw-ahbla-11,bdsssw-chgds-10,dw-fcrn-13}, but also for extended queries~(cf.~Section~\ref{ssec:extensions}) and applications~(cf.~Section~\ref{ssec:applications}).

\paragraph{Reach.}

An earlier hierarchical approach is~\emph{Reach}~\cite{g-rbran-04}. Reach is a centrality measure on vertices. Let~$\apath$ be a shortest~$\sourcevertex$--$\targetvertex$ path that contains vertex~$\vertex$. The reach~$\reach(\vertex,\apath)$ of~$\vertex$ with respect to~$\apath$ is defined as~$\min\{\dist(\sourcevertex,\vertex),\dist(\vertex,\targetvertex)\}$. The~(global) reach of~$\vertex$ in the graph~$\graph$ is the maximum reach of~$\vertex$ over \emph{all} shortest paths that contain~$\vertex$. Like other centrality measures~\cite{be-namf-05}, reach captures the importance of vertices in the graph, with the advantage that it can be used to prune a Dijkstra-based search.

A reach-based~$\sourcevertex$--$\targetvertex$ query runs Dijkstra's algorithm, but prunes the search at any vertex~$\vertex$ for which both~$\dist(\sourcevertex,\vertex) > \reach(\vertex)$ and~$\dist(\vertex,\targetvertex) > \reach(\vertex)$ hold; the shortest~$\sourcevertex$--$\targetvertex$ path provably does not contain~$\vertex$. To check these conditions, it suffices~\cite{gkw-raspa-09} to run bidirectional searches, each using the radius of the opposite search as a lower bound on~$\dist(\vertex,\targetvertex)$~(during the forward search) or~$\dist(\sourcevertex, \vertex)$~(backward search).

Reach values are determined during the preprocessing stage. Computing exact reaches requires computing shortest paths for all pairs of vertices, which is too expensive on large road networks. But the query is still correct if~$\reach(\vertex)$ represents only an upper bound on the reach of~$\vertex$. Gutman~\cite{g-rbran-04} has shown that such bounds can be obtained faster by computing partial shortest path trees. Goldberg~et~al.~\cite{gkw-raspa-09} have shown that adding shortcuts to the graph effectively reduces the reaches of most vertices, drastically speeding up both queries and preprocessing and making the algorithm practical for continent-sized networks.

\subsection{Bounded-Hop Techniques}

The idea behind bounded-hop techniques is to precompute distances between pairs of vertices, implicitly adding ``virtual shortcuts'' to the graph. Queries can then return the length of a virtual path with very few hops. Furthermore, they use only the precomputed distances between pairs of vertices, and not the input graph. A na\"ive approach is to use single-hop paths, i.\,e., precompute the distances among \emph{all} pairs of vertices~$\vertexa,\vertexb\in\vertices$. A single table lookup then suffices to retrieve the shortest distance. While the recent PHAST algorithm~\cite{dgnw-phast-13} has made precomputing all-pairs shortest paths feasible, storing all~$\Theta(\abs{\vertices}^2)$ distances is prohibitive already for medium-sized road networks. As we will see in this section, considering paths with slightly more hops~(two or three) leads to algorithms with much more reasonable trade-offs.

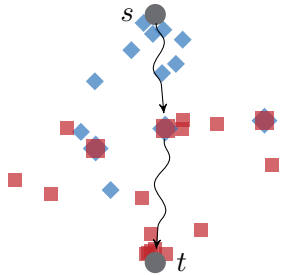
\begin{wrapfigure}{O}{0pt} \begin{tikzpicture}[scale=.66] \newcommand{\drawhubs}[4]{ \foreach \x in {1,...,#1} { \pgfmathsetmacro{\radi}{(sqrt(4)*rnd)^2} \node[smallvertex,#3,fill=#2,minimum size=#4{}pt,fill opacity=.7] at (90-60+120*rnd:\radi+.25) {}; } } \begin{scope}[yshift=5cm,rotate=180] \drawhubs{10}{kit-blau}{diamond}{7} \end{scope} \drawhubs{15}{kit-rot}{rectangle}{5} \node[smallvertex] (S) at (0,5) [label=left:$\sourcevertex$] {}; \node[smallvertex] (T) at (0,0) [label=right:$\targetvertex$] {}; \node[smallvertex,diamond,minimum size=10pt,fill=kit-blau,fill opacity=.7] (I1a) at (-1.2, 2.3) {}; \node[smallvertex,rectangle,minimum size=7pt,fill=kit-rot,fill opacity=.7] (I1b) at (-1.2, 2.3) {}; \node[smallvertex,diamond,minimum size=10pt,fill=kit-blau,fill opacity=.7] (I2a) at (.2, 2.7) {}; \node[smallvertex,rectangle,minimum size=7pt,fill=kit-rot,fill opacity=.7] (I2b) at (.2, 2.7) {}; \node[smallvertex,diamond,minimum size=10pt,fill=kit-blau,fill opacity=.7] (I3a) at (2.2, 2.86) {}; \node[smallvertex,rectangle,minimum size=7pt,fill=kit-rot,fill opacity=.7] (I3b) at (2.2, 2.86) {}; \path[squiggly arc] (S) -- (I2a); \path[squiggly arc] (I2a) -- (T); \end{tikzpicture} \caption[Illustration of a Hub Labels query]{Illustrating hub labels of vertices~$\sourcevertex$~(diamonds) and~$\targetvertex$~(squares).} \label{fig:literature:hublabels} \end{wrapfigure}

\paragraph{Labeling Algorithms.} We first consider \emph{labeling algorithms}~\cite{p-ppls-00}. During preprocessing, a \emph{label}~$\lab(\vertex)$ is computed for each vertex~$\vertex$ of the graph, such that, for any pair~$\vertexa,\vertexb$ of vertices, the distance~$\dist(\vertexa,\vertexb)$ can be determined by only looking at the labels~$\lab(\vertexa)$ and~$\lab(\vertexb)$. A natural special case of this approach is \emph{Hub Labeling}~(HL)~\cite{gppr-dlg-04,chkz-rdqhl-03}, in which the label~$\lab(\vertex)$ associated with vertex~$\vertex$ consists of a set of vertices~(the \emph{hubs} of~$\vertex$), together with their distances from~$\vertex$. These labels are chosen such that they obey the \emph{cover property}: for any pair~$(\sourcevertex,\targetvertex)$ of vertices,~$\lab(\sourcevertex) \cap \lab(\targetvertex)$ must contain at least one vertex on the shortest~$\sourcevertex$--$\targetvertex$ path. Then, the distance~$\dist(\sourcevertex,\targetvertex)$ can be determined in linear~(in the label size) time by evaluating~$\dist(\sourcevertex,\targetvertex) = \min\{\dist(\sourcevertex,\vertex) + \dist(\vertex,\targetvertex) \mid \vertex \in \lab(\sourcevertex)\ \text{and}\ \vertex \in \lab(\targetvertex)\}$. See ~\figurename~\ref{fig:literature:hublabels} for an illustration. For directed graphs, the label associated with~$\vertex$ is actually split in two: the forward label~$\lab_f(\vertex)$ has distances from~$\vertex$ \emph{to} the hubs, while the backward label~$\lab_b(\vertex)$ has distances \emph{from} the hubs to~$\vertex$; the shortest~$\sourcevertex$--$\targetvertex$ path has a hub in~$\lab_f(\sourcevertex) \cap \lab_b(\targetvertex)$.

Although the required average label size can be~$\Theta(\abs{\vertices})$ in general~\cite{gppr-dlg-04}, it can be significantly smaller for some graph classes. For road networks, Abraham et al.~\cite{adgw-ahbla-11} have shown that one can obtain good results by defining the label of vertex~$\vertex$ as the~(upward) search space of a CH query from~$\vertex$~(with suboptimal entries removed). In general, any vertex ordering fully defines a labeling~\cite{adgw-hhlsp-12}, and an ordering can be converted into the corresponding labeling efficiently~\cite{adgw-hhlsp-12,aiy-f-13}. The CH-induced order works well for road networks. For even smaller labels, one can pick the most important vertices greedily, based on how many shortest paths they hit~\cite{adgw-hhlsp-12}. A sampling version of this greedy approach works efficiently for a wide range of graph classes~\cite{dgpw-rdqmn-14}.

Note that, if labels are sorted by hub ID, a query consists of a linear sweep over two arrays, as in mergesort. Not only is this approach very simple, but it also has an almost perfect locality of access. With careful engineering, one does not even have to look at all the hubs in a label \cite{adgw-ahbla-11}. As a result, HL has the fastest known queries for road networks, taking roughly the time needed for five accesses to main memory~(see Section~\ref{sec:experiments}). One drawback is space usage, which, although not prohibitive, is significantly higher than for competing methods. By combining common substructures that appear in multiple labels, \emph{Hub Label Compression}~(HLC)~\cite{dgw-hlc-13} (see also~\cite{dgpw-rdqmn-14}) reduces space usage by an order of magnitude, at the expense of higher query times.

\paragraph{Transit Node Routing.}

The~\emph{Transit Node Routing}~(TNR)~\cite{bfss-frrnt-07,bfm-uspqt-09,ss-racts-09,als-tnrr-13} technique uses distance tables on a subset of the vertices. During preprocessing, it selects a small set~$\transitnodes \subseteq \vertices$ of \emph{transit nodes} and computes all pairwise distances between them. From those, it computes, for each vertex~$\vertex\in\vertices\setminus\transitnodes$, a relevant set of \emph{access nodes}~$\accessnodes(\vertex) \subseteq \transitnodes$. A transit node~$\vertexb \in \transitnodes$ is an access node of~$\vertexa$ if there is a shortest path~$\apath$ from~$\vertexa$ in~$\graph$ such that~$\vertexb$ is the first transit node contained in~$\apath$. In addition to the vertex itself, preprocessing also stores the distances between~$\vertexa$ and its access nodes.

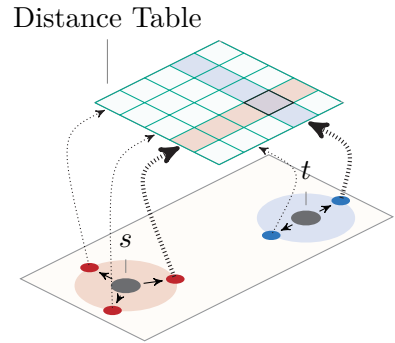
\begin{wrapfigure}{O}{0pt} \begin{tikzpicture}[scale=.66] \def\sly{.5} \def\slx{-1} \def\shy{2.8cm} \begin{scope}[yslant=\sly,xslant=\slx,every node/.append style={ yslant=\sly,xslant=\slx}] \draw[kit-schwarz!50, fill=kit-orange!2.5, fill opacity=.75] (0,0) rectangle (5,2.5); \begin{scope}[xshift=.92cm,yshift=1.3cm,every node/.append style={inner sep=0pt,circle,fill=kit-rot,minimum size=5pt}] \path[fill=kit-rot!15] (0,0) circle (.73cm); \coordinate (S1) at (90:.73); \node (NS1) at (S1) {}; \coordinate (S2) at (210:.73); \node (NS2) at (S2) {}; \coordinate (S3) at (330:.73); \node (NS3) at (S3) {}; \node[smallvertex] (S) at (0,0) {}; \draw[->] (S) edge (NS1) edge (NS2) edge (NS3); \end{scope} \begin{scope}[xshift=4.1cm, yshift=0.85cm,every node/.append style={inner sep=0pt,circle,fill=kit-blau,minimum size=5pt}] \path[fill=kit-blau!15] (0,0) circle (.7cm); \coordinate (T1) at at (0:.7cm); \node (NT1) at (T1) {}; \coordinate (T2) at (180:.7cm) {}; \node (NT2) at (T2) {}; \node[smallvertex] (T) at (0,0) {}; \draw[->] (T) edge (NT1) edge (NT2); \end{scope} \end{scope} \begin{scope}[yshift=\shy,yslant=\sly,xslant=\slx,every node/.append style={ yslant=\sly,xslant=\slx}] \end{scope} \begin{scope}[yshift=\shy,yslant=\sly,xslant=\slx,every node/.append style={yslant=\sly,xslant=\slx}] \begin{scope}[xshift=1.5cm] \draw[kit-schwarz!50, fill=kit-gruen!2.5, fill opacity=.75] (0,0) rectangle (2.5,2.5); \path[fill=kit-rot!30, fill opacity=.5] (0,.5) rectangle (2.5,1); \path[fill=kit-blau!30, fill opacity=.5] (1.5,0) rectangle (2,2.5); \draw[step=5mm, kit-gruen] (0,0) grid (2.5,2.5); \draw[kit-schwarz] (1.5,0.5) rectangle (2,1); \coordinate (DS1) at (0,2.25); \coordinate (DS1C) at (-1,2.25); \coordinate (DS2) at (0,1.25); \coordinate (DS2C) at (-1,1.25); \coordinate (DS3) at (0,0.75); \coordinate (DS3C) at (-1,0.75); \coordinate (DT1) at (1.75,0); \coordinate (DT1C) at (1.75,-1); \coordinate (DT2) at (0.75,0); \coordinate (DT2C) at (0.75,-1); \coordinate (DP) at (0.25,2.5); \end{scope} \end{scope} \begin{scope}[->,kit-schwarz,densely dotted] \draw (S1) .. controls (S1) and (DS1C) .. (DS1); \draw (S2) .. controls (S2) and (DS2C) .. (DS2); \draw[ultra thick] (S3) .. controls (S3) and (DS3C) .. (DS3); \draw[ultra thick] (T1) .. controls (T1) and (DT1C) .. (DT1); \draw (T2) .. controls (T2) and (DT2C) .. (DT2); \end{scope} \node [pin={[pin distance=.66cm]90:Distance Table}] at (DP) {}; \node [pin=90:$\sourcevertex$] at (S) {}; \node [pin=90:$\targetvertex$] at (T) {}; \end{tikzpicture} \caption[Illustration of a Transit Node Routing query]{Illustrating a TNR query. The access nodes of~$\sourcevertex$~($\targetvertex$) are indicated by three~(two) dots. The arrows point to the respective rows/columns of the distance table. The highlighted entries correspond to the access nodes which minimize the combined~$\sourcevertex$--$\targetvertex$ distance.} \label{fig:literature:tnr} \end{wrapfigure}

An~$\sourcevertex$--$\targetvertex$ query uses the distance table to select the path that minimizes the combined~$\sourcevertex$--$a(\sourcevertex)$--$a(\targetvertex)$--$\targetvertex$ distance, where~$a(\sourcevertex) \in \accessnodes(\sourcevertex)$ and~$a(\targetvertex) \in \accessnodes(\targetvertex)$ are access nodes. Note that the result is incorrect if the shortest path does not contain a vertex from~$\transitnodes$. To account for such cases, a \emph{locality filter} decides whether the query might be local~(i.\,e., does not contain a vertex from~$\transitnodes$). In that case, a fallback shortest path algorithm~(typically CH) is run to compute the correct distance. Note that TNR is still correct even if the locality filter occasionally misclassifies a global query as local. See~\figurename~\ref{fig:literature:tnr} for an illustration of a TNR query. Interestingly, global TNR queries~(which use the distance tables) tend to be faster than local ones~(which perform graph searches). To accelerate local queries, TNR can be extended to multiple~(hierarchical) layers of transit~(and access) nodes~\cite{bfm-uspqt-09,ss-racts-09}.

The choice of the transit node set is crucial to the performance of the algorithm. A natural approach is to select vertex separators or boundary vertices of arc separators as transit nodes. In particular, using grid-based separators yields natural locality filters and works well enough in practice for road networks~\cite{bfm-uspqt-09}. (Although an optimized preprocessing routine for this grid-based approach was later shown to have a flaw that could potentially result in suboptimal queries~\cite{wxdczz-spdqr-12}, the version with slower preprocessing reported in~\cite{bfm-uspqt-09} is correct and achieves the same query times.)

For better performance~\cite{ss-racts-09,gssv-erlrn-12,als-tnrr-13,ADFGW-13}, one can pick as transit nodes vertices that are classified as important by a hierarchical speedup technique~(such as CH). Locality filters are less straightforward in such cases: although one can still use geographical distances~\cite{ss-racts-09,gssv-erlrn-12}, a graph-based approach considering the Voronoi regions~\cite{m-as-88} induced by transit nodes tends to be significantly more accurate~\cite{als-tnrr-13}. A theoretically justified TNR variant~\cite{ADFGW-13} also picks important vertices as transit nodes and has a natural graph-based locality filter, but is impractical for large networks.

\paragraph{Pruned Highway Labeling.} The Pruned Highway Labeling~(PHL)~\cite{aikk-fspdq-14} algorithm can be seen as a hybrid between pure labeling and transit nodes. Its preprocessing routine decomposes the input into disjoint shortest paths, then computes a label for each vertex~$\vertexb$ containing the distance from~$\vertexb$ to vertices in a small subset of such paths. The labels are such that any shortest~$\sourcevertex$--$\targetvertex$ path can be expressed as~$\sourcevertex$--$\vertexa$--$\vertexc$--$\targetvertex$, where~$\vertexa$--$\vertexc$ is a subpath of a path~$\apath$ that belongs to the labels of~$\sourcevertex$ and~$\targetvertex$. Queries are thus similar to HL, finding the lowest-cost intersecting path. For efficient preprocessing, the algorithm uses the pruned labeling technique~\cite{aiy-f-13}. Although this method has some similarity with Thorup's distance oracle for planar graphs~\cite{t-corad-04}, it does not require planarity. PHL has only been evaluated on undirected graphs, however.

\subsection{Combinations}

Since the individual techniques described so far exploit different graph properties, they can often be combined for additional speedups. This section describes such hybrid algorithms. In particular, early results~\cite{sww-daola-00,hsww-cstsp-06} considered the combination of Geometric Containers, multilevel overlay graphs, and~(Euclidean-based) \astar on transportation networks, resulting in speedups of one or two orders of magnitude over Dijkstra's algorithm.

More recent studies have focused on combining hierarchical methods~(such as CH or Reach) with fast goal-directed techniques~(such as ALT or Arc Flags). For instance, the \emph{REAL} algorithm combines Reach and ALT~\cite{gkw-raspa-09}. A basic combination is straightforward: one simply runs an ALT query with additional pruning by reach~(using the ALT lower bounds themselves for reach evaluations). A more sophisticated variant uses \emph{reach-aware landmarks}: landmarks and their distances are only precomputed for vertices with high reach values. This saves space~(only a small fraction of the graph needs to store landmark distances), but requires two-stage queries~(goal direction is only used when the search is far enough from both source and target).

A similar space-saving approach is used by \emph{Core-ALT}~\cite{bdsssw-chgds-10,dn-crdtd-12}. It first computes an overlay graph for the \emph{core graph}, a~(small) subset~(e.\,g., 1\,\%) of vertices~(which remain after ``unimportant'' ones are contracted), then computes landmarks for the core vertices only. Queries then work in two stages: first plain bidirectional search, then ALT is applied when the search is restricted to the core. The~(earlier)~\emph{HH*} approach~\cite{dssw-hhs-09} is similar, but uses Highway Hierarchies~\cite{ss-ehh-12} to determine the core.

Another approach with two-phase queries is \emph{ReachFlags}~\cite{bdsssw-chgds-10}. During preprocessing, it first computes~(approximate) reach values for all vertices in~$\graph$, then extracts the subgraph~$H$ induced by all vertices whose reach value exceeds a certain threshold. Arc flags are then only computed for~$H$, to be used in the second phase of the query.

The \emph{SHARC} algorithm~\cite{bd-sharc-09} combines the computation of shortcuts with multilevel arc flags. The preprocessing algorithm first determines a partition of the graph and then computes shortcuts and arc flags in turn. Shortcuts are obtained by contracting unimportant vertices with the restriction that shortcuts never span different cells of the partition. The algorithm then computes arc flags such that, for each cell~$\cell$, the query uses a shortcut arc if and only if the target vertex is not in~$\cell$. Space usage can be reduced with various compression techniques~\cite{bdgw-sesr-10}. Note that SHARC is unidirectional and hierarchical: arc flags not only guide the search toward the target, but also vertically across the hierarchy. This is useful when the backward search is not well defined, as in time-dependent route planning~(discussed in Section~\ref{ssec:extensions}).

Combining CH with Arc Flags results in the \emph{CHASE} algorithm~\cite{bdsssw-chgds-10}. During preprocessing, a regular contraction hierarchy is computed and the search graph that includes all shortcuts is assembled. The algorithm then extracts the subgraph~$H$ induced by the top~$k$ vertices according to the contraction order. Bidirectional arc flags~(and the partition) are finally computed on the restricted subgraph~$H$. Queries then run in two phases. Since computing arc flags was somewhat slow,~$k$ was originally set to a small fraction~(about~5\,\%) of the total number~$\abs{\vertices}$ of vertices~\cite{bdsssw-chgds-10}. More recently, Delling et al.~showed that PHAST~(see Section~\ref{ssec:extensions}) can compute arc flags fast enough to allow~$k$ to be set to~$\abs{\vertices}$, making CHASE queries much simpler~(single-pass), as well as faster~\cite{dgnw-phast-13}.

Finally, Bauer et al.~\cite{bdsssw-chgds-10} combine Transit Node Routing with Arc Flags to obtain the TNR+AF algorithm. Recall that the bottleneck of the TNR query is performing the table lookups between pairs of access nodes from~$\accessnodes(\sourcevertex)$ and~$\accessnodes(\targetvertex)$. To reduce the number of lookups, TNR+AF's preprocessing decomposes the set of transit nodes~$\transitnodes$ into~$k$ cells. For each vertex~$\sourcevertex$ and access node~$\vertexa\in\accessnodes(\sourcevertex)$, it stores a~$k$-bit vector, with bit~$i$ indicating whether there exists a shortest path from~$\sourcevertex$ to cell~$i$ through~$\vertexa$. A query then only considers the access nodes from~$\sourcevertex$ that have their bits set with respect to the cells of~$\accessnodes(\targetvertex)$. A similar pruning is done at the target.

\subsection{Extensions}\label{ssec:extensions}

In various applications, one is often interested in more than just the length of the shortest path between two points in a static network. Most importantly, one should also be able to retrieve the shortest path itself. Moreover, many of the techniques considered so far can be adapted to compute batched shortest paths~(such as distance tables), to more realistic scenarios~(such as dynamic networks), or to deal with multiple objective functions. In the following, we briefly discuss each of these extensions.

\subsubsection{Path Retrieval} Our descriptions so far have focused on finding only the \emph{length} of the shortest path. The algorithms we described can easily be augmented to provide the actual list of edges or vertices on the path. For techniques that do not use shortcuts~(such as Dijkstra's algorithm, \astar search, or Arc Flags), one can simply maintain a parent pointer for each vertex~$\vertexb$, updating it whenever the distance label of~$\vertexb$ changes. When shortcuts are present~(such as in CH, SHARC, or CRP), this approach gives only a \emph{compact} representation of the shortest path~(in terms of shortcuts). The shortcuts then need to be unpacked. If each shortcut is the concatenation of two other arcs~(or shortcuts), as in CH, storing the middle vertex~\cite{gssv-erlrn-12} of each shortcut allows for an efficient~(linear-time) recursive unpacking of all shortcuts on the output path. If shortcuts are built from multiple arcs~(as for CRP or SHARC), one can either store the entire sequence for each shortcut~\cite{ss-ehh-12} or run a local~(bidirectional) Dijkstra search from its endpoints~\cite{dgpw-crprn-13}. These two techniques can be used for bounded-hop algorithms as well.

\subsubsection{Batched Shortest Paths} \label{ssec:batched}

Some applications require computing multiple paths at once. For example, advanced logistics applications may need to compute all distances between a source set~$S$ and a target set~$T$. This can be trivially done with~$\abs{S} \cdot \abs{T}$ point-to-point shortest-path computations. Using a hierarchical speedup technique~(such as CH), this can be done in time comparable to~$\bigO(\abs{S}+\abs{T})$ point-to-point queries in practice, which is much faster. First, one runs a backward upward search from each~$\targetvertex_i \in T$; for each vertex~$\vertex$ scanned during the search from~$\targetvertex_i$, one stores its distance label~$d_{\targetvertex_i}(\vertex)$ in a bucket~$\beta(\vertex)$. Then, one runs a forward upward search from each~$\sourcevertex_j \in S$. Whenever such a search scans a vertex~$\vertexb$ with a non-empty bucket, one searches the bucket and checks whether~$d_{\sourcevertex_j}(\vertexb) + d_{\targetvertex_i}(\vertexb)$ improves the best distance seen so far between~$\sourcevertex_j$ and~$\targetvertex_i$. This \emph{bucket-based approach} was introduced for Highway Hierarchies~\cite{ksssw-cmmsp-07}, but can be used with any other hierarchical speedup technique~(such as CH) and even with hub labels~\cite{dgw-fbspr-11}. When the bucket-based approach is combined with a separator-based technique~(such as CRP), it is enough to keep buckets only for the boundary vertices~\cite{dw-cpiqr-13}. Note that this approach can be used to compute one-to-many or many-to-many distances.

Some applications require one-to-all computations, i.\,e., finding the distances from a source vertex~$\sourcevertex$ to all other vertices in the graph. For this problem, Dijkstra's algorithm is optimal in the sense that it visits each edge exactly once, and hence runs in essentially linear time~\cite{g-apspa-08}. However, Dijkstra's algorithm has bad locality and is hard to parallelize, especially for sparse graphs~\cite{ms-dsaps-03,mbbc-pspas-09}. PHAST~\cite{dgnw-phast-13} builds on CH to improve this. The idea is to split the search in two phases. The first is a forward upward search from~$\sourcevertex$, and the second runs a linear scan over the shortcut-enriched graph, with distance values propagated from more to less important vertices. Since the instruction flow of the second phase is~(almost) independent of the source, it can be engineered to exploit parallelism and improve locality. In road networks, PHAST can be more than an order of magnitude faster than Dijkstra's algorithm, even if run sequentially, and can be further accelerated using multiple cores and even GPUs. This approach can also be extended to the \emph{one-to-many problem}, i.\,e., computing distances from a source to a subset of predefined targets~\cite{dgw-fbspr-11}. Similar techniques can also be applied with graph separators~(instead of CH), yielding comparable query times but with faster~(metric-dependent) preprocessing~\cite{ep-grasp-14}.

\subsubsection{Dynamic Networks}

Transportation networks tend to be dynamic, with unpredictable delays, traffic, or closures. If one assumes that the modified network is stable for the foreseeable future, the obvious approach for speedup techniques to deal with this is to rerun the preprocessing algorithm. Although this ensures queries are as fast as in the static scenario, it can be quite costly. As a result, four other approaches have been considered.

It is often possible to just ``repair'' the preprocessed data instead of rebuilding it from scratch. This approach has been tried for various techniques, including Geometric Containers~\cite{wwz-gcesp-05}, ALT~\cite{dw-lbrdg-07}, Arc Flags~\cite{ddfv-fdmaf-12}, and CH~\cite{ss-dhnr-07,gssv-erlrn-12}, with varying degrees of success. For CH, for example, one must keep track of dependencies between shortcuts, partially rerunning the contraction as needed. Changes that affect less important vertices can be dealt with faster.

Another approach is to adapt the query algorithms to work around the ``wrong'' parts of the preprocessing phase. In particular, ALT is resilient to increases in arc costs~(due to traffic, for example): queries remain correct with the original preprocessing, though query times may increase~\cite{dw-lbrdg-07}. Less trivially, CH queries can also be modified to deal with dynamic changes to the network~\cite{ss-dhnr-07,gssv-erlrn-12} by allowing the search to bypass affected shortcuts by going ``down'' the hierarchy. This is useful when queries are infrequent relative to updates.

A third approach is to make the preprocessing stage completely metric-independent, shifting all metric-dependent work to the query phase. Funke et al.~\cite{fns-opca-14} generalize the multilevel overlay graph approach to encode \emph{all}~$k$-hop paths~(for small~$k$) in an overlay graph. Under the assumption that edge costs are defined by a small number of physical parameters~(as in simplified road networks) this allows setting the edge costs at query time, though queries become significantly slower.

For more practical queries, the fourth approach splits the preprocessing phase into metric-independent and metric-dependent stages. The metric-independent phase takes as input only the network topology, which is fairly stable. When edge costs change~(which happens often), only the~(much cheaper) metric-dependent stage must be rerun, partially or in full. This concept can again be used for various techniques, with ALT, CH, and CRP being the most prominent. For ALT, one can keep the landmarks, and just recompute the distances to them~\cite{dw-lbrdg-07,ep-olbrp-13}. For CH, one can keep the ordering, and just rerun contraction~\cite{gssv-erlrn-12,dsw-cch-sea-14}. For CRP, one can keep the partitioning and the overlay topology, and just recompute the shortcut lengths using a combination of contraction and graph searches~\cite{dgpw-crprn-13}. Since the contraction is metric-independent, one can precompute and store the sequence of contraction operations and reexecute them efficiently whenever edge lengths change~\cite{dgpw-crprn-13,dkw-cddgp-14}. The same approach can be used for CH with metric-independent orders~\cite{dsw-cch-sea-14}.

\subsubsection{Time-Dependence} In real transportation networks, the best route often depends on the departure time in a predictable way \cite{dbs-a-10}. For example, certain roads are consistently congested during rush hours, and certain buses or trains run with different frequencies during the day. When one is interested in the earliest possible arrival given a specified departure time~(or, symmetrically, the latest departure), one can model this as the \emph{time-dependent} shortest path problem, which assigns travel time functions to~(some of) the edges, representing how long it takes to traverse them at each time of the day. Dijkstra's algorithm still works~\cite{ch-tsrtn-66} as long as later departures cannot lead to earlier arrivals; this \emph{non-overtaking} property is often called first-in first-out~(FIFO). Although one must deal with functions instead of scalars, the theoretical running time of Dijkstra-based algorithms can still be bounded~\cite{dos-sptdf-12,fhs-octds-14}. Moreover, many of the techniques described so far work in this scenario, including bidirectional ALT~\cite{ndls-bastd-12,dn-crdtd-12}, CH~\cite{bgsv-mtdtt-13}, or SHARC~\cite{d-tdsr-11}. Recently, Kontogiannis and Zaroliagis~\cite{kz-dotdn-14} have introduced a theoretical~(approximate) distance oracle with sublinear running time. Other scenarios~(besides FIFO with no waiting at vertices) have been studied~\cite{or-spmda-90,or-mwptd-91,d-amcpt-04,d-spfif-04}, but they are less relevant for transportation networks.

There are some challenges, however. In particular, bidirectional search becomes more complicated~(since the time of arrival is not known), requiring changes to the backward search~\cite{ndls-bastd-12,bgsv-mtdtt-13}. Another challenge is that shortcuts become more space-consuming~(they must model a more complicated travel time function), motivating compression techniques that do not sacrifice correctness, as demonstrated for SHARC~\cite{bdgw-sesr-10} or CH~\cite{bgsv-mtdtt-13}. Batched shortest paths can be computed in such networks efficiently as well~\cite{gs-etdmm-10}.

Time-dependent networks motivate some elaborate~(but still natural) queries, such as finding the best departure time in order to minimize the total time in transit. Such queries can be dealt with by~\emph{range searches}, which compute the travel time function between two points. There exist Dijkstra-based algorithms~\cite{dos-sptdf-12} for this problem, and most speedup techniques can be adapted to deal with this as well~\cite{d-tdsr-11,bgsv-mtdtt-13}.

Unfortunately, even a slight deviation from the travel time model, where total cost is a linear combination of travel time and a constant cost offset, makes the problem NP-hard~\cite{aops-dspmt-03,bs-tdrpg-12}. However, a heuristic adaptation of time-dependent CH shows negligible errors in practice~\cite{bs-tdrpg-12}.

\subsubsection{Multiple Objective Functions} Another natural extension is to consider multiple cost functions. For example, certain vehicle types cannot use all segments of the transportation network. One can either adapt the preprocessing such that these \emph{edge restrictions} can be applied during query time~\cite{grst-rpfer-12}, or perform a metric update for each vehicle type.

Also, the search request can be more flexible. For example, one may be willing to take a more scenic route even if the trip is slightly longer. This can be dealt with by performing a multicriteria search. In such a search, two paths are incomparable if neither is better than the other in all criteria. The goal is to find a \emph{Pareto set}, i.\,e., a maximum set of incomparable paths. Such sets of shortest paths can be computed by extensions of Dijkstra's algorithm; see~\cite{eg-mcosa-02} for a survey on multicriteria combinatorial optimization. More specifically, the \emph{Multicriteria Label-Setting}~(MLS) algorithm~\cite{h-bpp-79,m-omspp-84,t-rmlw-95,m-vvpgm-99} extends Dijkstra's algorithm by keeping, for each vertex, a \emph{bag} of nondominated labels. Each label is represented as a tuple, with one entry per optimization criterion. The priority queue maintains labels instead of vertices, typically ordered lexicographically. In each iteration, it extracts the minimum label~$\lab$ and scans the incident arcs~$\arc = (\vertexa,\vertexb)$ of the vertex~$\vertexa$ associated with~$\lab$. It does so by adding the cost of~$\arc$ to~$\lab$ and then merging~$\lab$ into the bag of~$\vertexb$, eliminating possibly dominated labels on the fly. In contrast, the \emph{Multi-Label-Correcting}~(MLC) algorithm~\cite{d-ctdsp-99,dw-tdrp-09} considers the whole bag of nondominated labels associated with~$\vertexa$ at once when scanning the vertex~$\vertexa$. Hence, individual labels of~$\vertexa$ may be scanned multiple times during one execution of the algorithm.

Both MLS and MLC are fast enough as long as the Pareto sets are small~\cite{mw-pspof-01,dms-mcspt-08}. Unfortunately, Pareto sets may contain exponentially many solutions, even for the restricted case of two optimization criteria~\cite{h-bpp-79}, which makes it hard to achieve large speedups~\cite{dw-pps-09,bdgm-atdmc-09}. To reduce the size of Pareto sets, one can relax domination. In particular,~$(1+\varepsilon)$-Pareto sets have provable polynomial size~\cite{py-o-00} and can be computed efficiently~\cite{loridan1984,w-ee-86,tz-moifp-09}. Moreover, large Pareto sets open up a potential for parallelization that is not present for a single objective function~\cite{sm-plsmo-13,eks-pbspu-14}.

A reasonable alternative~\cite{gks-rpfof-10} to multicriteria search is to optimize a linear combination~$\alpha c_1 + (1-\alpha) c_2$ of two criteria~$(c_1,c_2)$, with the parameter~$\alpha$ set at query time. Moreover, it is possible to efficiently compute the values of~$\alpha$ where the path actually changes. Funke and Storandt~\cite{fs-pcchm-13} show that CH can handle such functions with polynomial preprocessing effort, even with more than two criteria.

\subsection{Theoretical Results}

Most of the algorithms mentioned so far were developed with practical performance in mind. Almost all methods we surveyed are exact: they provably find the exact shortest path. Their performance~(in terms of both preprocessing and queries), however, varies significantly with the input graph. Most algorithms work well for real road networks, but are hardly faster than Dijkstra's algorithm on some other graph classes. This section discusses theoretical work that helps understand why the algorithms perform well and what their limitations are.

Most of the algorithms considered have some degree of freedom during preprocessing~(such as which partition, which vertex order, or which landmarks to choose). An obvious question is whether one could efficiently determine the best such choices for a particular input so as to minimize the query search space~(a natural proxy for query times). Bauer et al.~\cite{bckkw-psuth-10} have determined that finding optimal landmarks for ALT is NP-hard. The same holds for Arc Flags~(with respect to the partition), SHARC~(with respect to the shortcuts), Multilevel Overlay Graphs~(with respect to the separator), Contraction Hierarchies~(with respect to the vertex order), and Hub Labels~(with respect to the hubs)~\cite{Weller14a}. In fact, minimizing the number of shortcuts for CH is APX-hard~\cite{bckkw-psuth-10,m-o-12}. For SHARC, however, a greedy factor-$k$ approximation algorithm exists~\cite{bddsw-tspca-12}. Deciding which~$k$ shortcuts~(for fixed~$k$) to add to a graph in order to minimize the SHARC search space is also NP-hard~\cite{bddsw-tspca-12}. Bauer et al.~\cite{bbrw-ocpga-13} also analyze the preprocessing of Arc Flags in more detail and on restricted graph classes, such as paths, trees, and cycles, and show that finding an optimal partition is NP-hard even for binary trees.

Besides complexity, theoretical performance bounds for query algorithms, which aim to explain their excellent practical performance, have also been considered. Proving better running time bounds than those of Dijkstra's algorithm is unlikely for general graphs; in fact, there are inputs for which most algorithms are ineffective. That said, one can prove nontrivial bounds for specific graph classes. In particular, various authors~\cite{m-o-12,bcrw-sssch-13} have independently observed a natural relationship between CH and the notions of filled graphs~\cite{p-tulgg-61} and elimination trees~\cite{s-anisg-82}. For planar graphs, one can use nested dissection~\cite{lrt-gnd-79} to build a CH order leading to~$\bigO(\abs{\vertices} \log \abs{\vertices})$ shortcuts~\cite{m-o-12,bcrw-sssch-13}. More generally, for minor-closed graph classes with balanced~$\bigO(\sqrt{\abs{\vertices}})$-separators, the search space is bounded by~$\bigO(\sqrt{\abs{\vertices}})$~\cite{bcrw-sssch-13}. Similarly, on graphs with treewidth~$k$, the search space of CH is bounded by~$\bigO(k \log \abs{\vertices})$~\cite{bcrw-sssch-13}.

Road networks have motivated a large amount of theoretical work on algorithms for planar graphs. In particular, it is known that planar graphs have separators of size~$\bigO(\sqrt{\abs{\vertices}})$~\cite{lt-astpg-79,lrt-gnd-79}. Although road networks are not strictly planar, they do have small separators~\cite{eg-s-08,dgrw-gpnc-11}, so theoretically efficient algorithms for planar graphs are likely to also perform well in road networks. Sommer~\cite{s-spqsn-14} surveys several approximate methods with various trade-offs. In practice, the observed performance of most speedup techniques is much better on actual road networks than on arbitrary planar graphs~(even grids). A theoretical explanation of this discrepancy thus requires a formalization of some property related to key features of real road networks.

One such graph property is~\emph{Highway Dimension}, proposed by Abraham et~al.~\cite{ADFGW-13}~(see also~\cite{adfgw-vcdsp-11,afgw-hdspp-10}). Roughly speaking, a graph has highway dimension~$h$ if, at any scale~$r$, one can hit all shortest paths of length at least~$r$ by a hitting set~$S$ that is \emph{locally sparse}, in the sense that any ball of radius~$r$ has at most~$h$ elements from~$S$. Based on previous experimental observations~\cite{bfss-frrnt-07}, the authors~\cite{afgw-hdspp-10} conjecture that road networks have small highway dimension. Based on this notion, they establish bounds on the performance of~(theoretically justified versions of) various speedup techniques in terms of~$h$ and the graph diameter~$D$, assuming the graph is undirected and that edge lengths are integral. More precisely, after running a polynomial-time preprocessing routine, which adds~$\bigO(h\log h\log D)$ shortcuts to~$\graph$, Reach and CH run in~$\bigO((h \log h \log D)^2)$ time. Moreover, they also show that HL runs in~$\bigO(h \log h \log D)$ time and long-range TNR queries take~$\bigO(h^2)$ time. In addition, Abraham et al.~\cite{ADFGW-13} show that a graph with highway dimension~$h$ has doubling dimension~$\log(h + 1)$, and Kleinberg et al.~\cite{ksw-teuss-04} show that landmark-based triangulation yields good bounds for most pairs of vertices of graphs with small doubling dimension. This gives insight into the good performance of ALT in road networks.

The notion of highway dimension is an interesting application of the scientific method. It was originally used to explain the good observed performance of CH, Reach, and TNR, and ended up predicting that HL~(which had not been implemented yet) would have good performance in practice.

Generative models for road networks have also been proposed and analyzed. Abraham et al.~\cite{ADFGW-13,afgw-hdspp-10} propose a model that captures some of the properties of road networks and generates graphs with provably small highway dimension. Bauer et al.~\cite{bkmw-srn-10} show experimentally that several speedup techniques are indeed effective on graphs generated according to this model, as well as according to a new model based on Voronoi diagrams. Models with a more geometric flavor have been proposed by Eppstein and Goodrich~\cite{eg-s-08} and by Eisenstat~\cite{e-rrntq-11}.

Besides these results, Rice and Tsotras~\cite{rt-basaa-12} analyze the~\astar algorithm and obtain bounds on the search space size that depend on the underestimation error of the potential function. Also, maintaining and updating multilevel overlay graphs have been theoretically analyzed in~\cite{bcddf-dmogs-08}. For Transit Node Routing, Eisner and Funke~\cite{ef-tnlbr-12} propose instance-based lower bounds on the size of the transit node set. For labeling algorithms, bounds on the label size for different graph classes are given by Gavoille et al.~\cite{gppr-dlg-04}. Approximation algorithms to compute small labels have also been studied~\cite{chkz-rdqhl-03,bggn-ahlo-13,dgsw-hltp-14}; although they can find slightly better labels than faster heuristics~\cite{adgw-hhlsp-12,dgpw-rdqmn-14}, their running time is prohibitive~\cite{dgsw-hltp-14}.

Because the focus of this work is on algorithm engineering, we refrain from going into more detail about the available theoretical work. Instead, we refer the interested reader to overview articles with a more theoretical emphasis, such as those by Sommer~\cite{s-spqsn-14}, Zwick~\cite{z-e-01}, and Gavoille and Peleg~\cite{gp-cldds-03}.

\section{Route Planning in Road Networks} \label{sec:road}

In this section, we experimentally evaluate how the techniques discussed so far perform in road networks. Moreover, we discuss applications of some of the techniques, as well as alternative settings such as databases or mobile devices.

\subsection{Experimental Results} \label{sec:experiments}

Our experimental analysis considers carefully engineered implementations, which is very important when comparing running times. They are written in C++ with custom-built data structures. Graphs are represented as adjacency arrays~\cite{ms-adstb-08}, and priority queues are typically binary heaps, 4-heaps, or multilevel buckets. As most arcs in road networks are bidirectional, state-of-the-art implementations use edge compression~\cite{s-rprn-08}: each road segment is stored at both of its endpoints, and each occurrence has two flags indicating whether the segment should be considered as an incoming and/or outgoing arc. This representation is compact and allows efficient iterations over incoming and outgoing arcs.

We give data for two models. The \emph{simplified model} ignores turn restrictions and penalties, while the \emph{realistic model} includes the turn information~\cite{w-mctrp-02}. There are two common approaches to deal with turns. The \emph{arc-based representation}~\cite{c-ofmrn-61} blows up the graph so that roads become vertices and feasible turns become arcs. In contrast, the \emph{compact representation}~\cite{gv-errnt-11,dgpw-crp-11} keeps intersections as vertices, but with associated \emph{turn tables}. One can save space by sharing turn tables among many vertices, since the number of intersection types in a road network is rather limited. Most speedup techniques can be used as is for the arc-based representation, but may need modification to work on the compact model.

Most experimental studies are restricted to the simplified model. Since some algorithms are more sensitive to how turns are modeled than others, it is hard to extrapolate these results to more realistic networks. We therefore consider experimental results for each model separately.

\subsubsection{Simplified Model} An important driving force behind the research on speedup techniques for Dijkstra's algorithm was its application to road networks. A key aspect for the success of this research effort was the availability of continent-sized benchmark instances. The most widely used instance has been the road network of Western Europe from PTV AG, with 18.0 million vertices and 42.5 million directed arcs. Besides ferries (for which the traversal time was given), it has 13 road categories. Category $i$ has been assigned an average speed of 10$i$\,km/h. This synthetic assignment is consistent with more realistic proprietary data~\cite{dgw-hlc-13,dgpw-crprn-13}. Another popular~(and slightly bigger) instance, representing the TIGER/USA road network, is undirected and misses several important road segments~\cite{adgw-arrn-13}. Although the inputs use the simplified model, they allowed researchers from various groups to run their algorithms on the same instance, comparing their performance. In particular, both instances were tested during the DIMACS Challenge on Shortest Paths~\cite{dgj-spndi-09}.

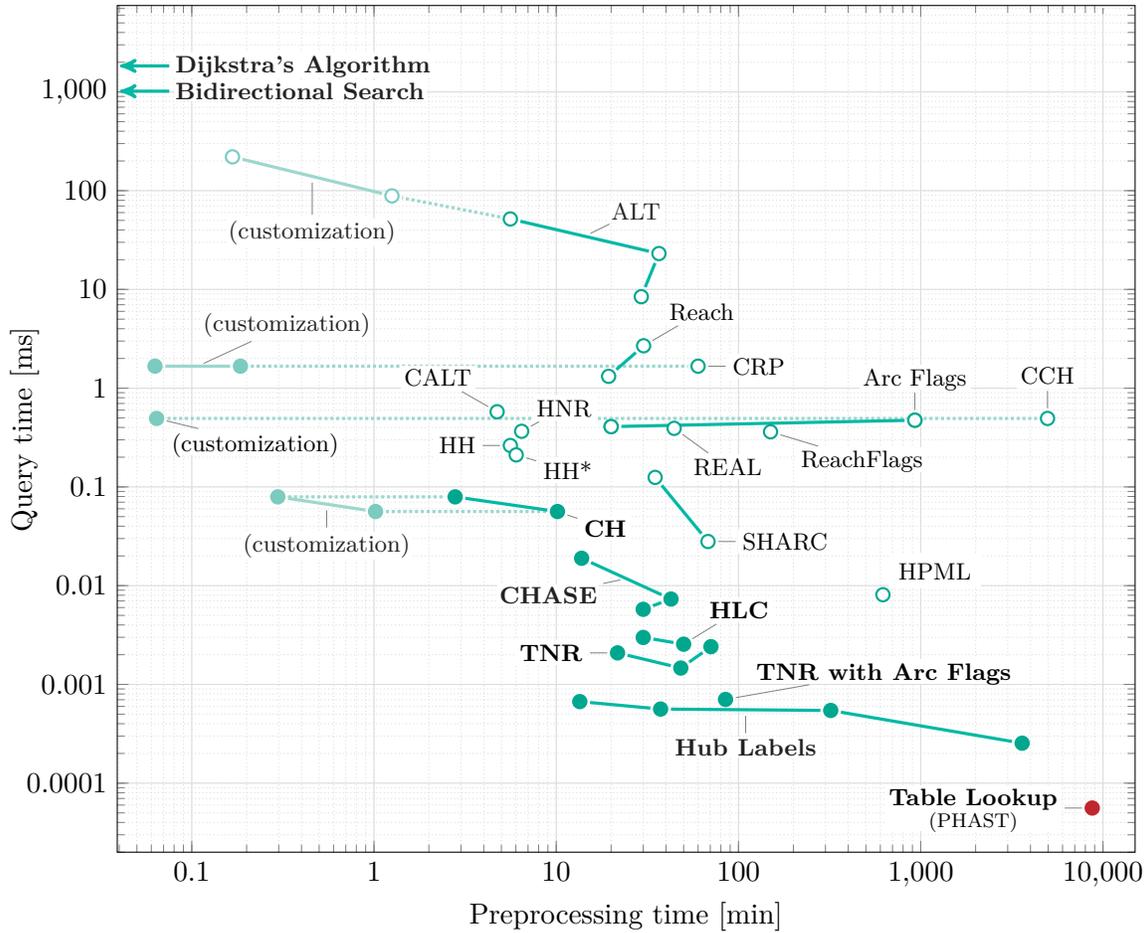
\begin{figure}[!t] \begin{tikzpicture} \begin{loglogaxis}[ width=\linewidth, height=.85\linewidth, log basis x=10, log basis y=10, xmin=0.039, xmax=15000, ymin=0.000020, ymax=7500, log ticks with fixed point, xlabel={Preprocessing time [min]}, ylabel={Query time [ms]}, grid=both, major grid style={color=kit-schwarz!15}, minor grid style={color=kit-schwarz!15,densely dotted}, ] \begin{scope}[every label/.style={text depth=0pt,fill=white,font=\smaller},every pin/.style={text depth=0pt,inner sep=.05cm,pin distance=.3cm,fill=white,font=\smaller}]

\node[plotpoint,pin={90:Arc Flags}] (AF) at (axis cs:\computeNine*2156,\computeNine*1.1) {}; \node[plotpoint,pin={180:HH}] (HH) at (axis cs:\computeNine*13,\computeNine*0.61) {}; \node[paretopoint] (TNR) at (axis cs:\computeNine*164,\computeNine*0.0056) {}; \node[paretopoint] (TNR2) at (axis cs:\computeNine*112,\computeNine*0.0034) {}; \node[plotpoint] (ALT-a16) at (axis cs:\computeNine*13,\computeNine*120.1) {}; \node[plotpoint] (ALT-m16) at (axis cs:\computeNine*85,\computeNine*53.6) {}; \node[plotpoint] (ALT-a64) at (axis cs:\computeNine*68,\computeNine*19.6) {}; \node[plotpoint,pin={[pin distance=.2cm]355:HH*}] (HH*) at (axis cs:\computeNine*14,\computeNine*0.49) {}; \node[plotpoint] (SHARC) at (axis cs:\computeNine*81,\computeNine*0.29) {}; \node[plotpoint,pin={0:SHARC}] (biSHARC) at (axis cs:\computeNine*158,\computeNine*0.065) {}; \node[paretopoint] (ecoCHASE) at (axis cs:\computeNine*32,\computeNine*0.044) {}; \node[paretopoint] (genCHASE) at (axis cs:\computeNine*99,\computeNine*0.017) {};

\node[paretopoint,pin={30:\textbf{TNR with Arc Flags}}] (TNRAF) at (axis cs:\chOpteron*229,\chOpteron*0.0019) {};

\node[plotpoint] (HPML) at (axis cs:\computeNine*1440,\computeNine*0.0188) [label=above right:HPML] {};

\node[paretopoint] (ecoCH) at (axis cs:\chOpteron*7.52,\chOpteron*0.214) {}; \node[paretopoint,pin={[pin distance=.2cm]350:\textbf{CH}}] (aggCH) at (axis cs:\chOpteron*27.4,\chOpteron*0.152) {}; \node[paretopoint,kit-gruen50] (ecoCustomCH) at (axis cs:\chOpteron*0.8,\chOpteron*0.214) {}; \node[paretopoint,kit-gruen50] (aggCustomCH) at (axis cs:\chOpteron*2.75,\chOpteron*0.152) {};

\node[plotpoint,pin={90:CCH}] (CCH) at (axis cs:\computeEleven*4151.42,\computeEleven*0.413) {}; \node[plotpoint,kit-gruen50,pin={[pin distance=.1cm]315:(customization)}] (customCCH) at (axis cs:\computeEleven*0.05367,\computeEleven*0.413) {};

\node[paretopoint,pin={180:\textbf{TNR}}] (CH-TNR) at (axis cs:\spaThree*34,\spaThree*0.00327) {};

\node[plotpoint,draw=kit-gruen50] (ALT-pfoser-8) at (axis cs:\pfoser*8/60,\pfoser*175.3) {}; \node[plotpoint,draw=kit-gruen50] (ALT-pfoser-64) at (axis cs:\pfoser*60/60,\pfoser*70.6) {};

\node[plotpoint] (PHASTAF) at (axis cs:\spaTwo*20,\spaTwo*0.408) {}; \node[paretopoint] (PHASTCHASE) at (axis cs:\spaTwo*30,\spaTwo*0.00576) {}; \node[paretopoint] (HL-17 local) at (axis cs:320.68,\spaTwo*0.000545) {}; \node[paretopoint] (HL-15 local) at (axis cs:\spaTwo*37.38,\spaTwo*0.000563) {}; \node[paretopoint] (HL-0 local) at (axis cs:13.46,\spaTwo*0.00067) {};

\node[paretopoint,pin={45:\textbf{HLC}}] (HLC-15) at (axis cs:\spaTwo*50,\spaTwo*0.002554) {}; \node[paretopoint] (HLC-0) at (axis cs:\spaTwo*30,\spaTwo*0.002989) {}; \node[paretopoint] (HL-inf global) at (axis cs:\spaTwo*3600,\spaTwo*0.000254) {};

\node[paretopoint,kit-gruen50] (CRP-5mc) at (axis cs:\spaTwo*0.06283,\spaTwo*1.67) {}; \node[paretopoint,kit-gruen50] (CRP-5) at (axis cs:\spaTwo*0.185,\spaTwo*1.67) {}; \node[plotpoint,pin={0:CRP}] (CRP-5 part) at (axis cs:\spaTwo*60,\spaTwo*1.67) {};

\node[paretopoint,kit-rot,pin={[align=center]180:{\textbf{Table Lookup}\\[-.75ex]\smaller{(PHAST)}}}] (PHAST) at (axis cs:\spaTwo*8730,\spaTwo*0.000056) {}; \end{scope}

\begin{scope}[-,very thick,kit-gruen!75,every node/.style={kit-schwarz,inner sep=.025cm,font=\smaller},every pin/.style={inner sep=.025cm,pin distance=.3cm,fill=white,font=\smaller}] \draw[<-] (axis cs:0.039,\spaTwo*1833.92) -- node [pos=1,fill=white,anchor=west] {\textbf{Dijkstra's Algorithm}} ++(0:.3); \draw[<-] (axis cs:0.039,\spaTwo*1017.16) -- node [pos=1,fill=white,anchor=west] {\textbf{Bidirectional Search}} ++(0:.3);

\draw[densely dotted,kit-gruen50!75] (ecoCustomCH) -- (ecoCH); \draw[densely dotted,kit-gruen50!75] (aggCustomCH) -- (aggCH); \draw[densely dotted,kit-gruen50!75] (CCH) -- (customCCH); \draw[kit-gruen50!75] (ecoCustomCH) -- node[fill=none,pos=.5,pin={270:(customization)}] {} (aggCustomCH); \draw[kit-gruen50!75] (CRP-5mc) -- node [fill=none,pos=.5,pin={85:(customization)}] {} (CRP-5); \draw[densely dotted,kit-gruen50!75] (CRP-5) -- (CRP-5 part); \draw (PHASTAF) -- (AF); \draw (TNR) -- (TNR2); \draw (TNR2) -- (CH-TNR); \draw (ALT-a16) -- node [pos=.5,pin={30:ALT}] {} (ALT-m16); \draw (ALT-m16) -- (ALT-a64); \draw[densely dotted,kit-gruen50!75] (ALT-a16) -- (ALT-pfoser-64); \draw[kit-gruen50!75] (ALT-pfoser-64) -- node [fill=none,pos=.5,pin={[pin distance=.5cm]270:(customization)}] {} (ALT-pfoser-8); \draw (SHARC) -- (biSHARC); \draw (ecoCHASE) -- node[fill=none,pos=.5,pin={190:\textbf{CHASE}}] {} (genCHASE); \draw (genCHASE) -- (PHASTCHASE); \draw (HL-0 local) -- (HL-15 local); \draw (HL-15 local) -- node[fill=none,pos=.5,pin={270:\textbf{Hub Labels}}] {} (HL-17 local); \draw (HL-17 local) -- (HL-inf global); \draw (HLC-0) -- (HLC-15); \draw (aggCH) -- (ecoCH); \end{scope}

\begin{scope}[every label/.style={text depth=0pt,fill=white,font=\smaller},every pin/.style={text depth=0pt,inner sep=.05cm,pin distance=.3cm,fill=white,font=\smaller}] \node[plotpoint,pin={45:Reach}] (RE) at (axis cs:\computeNine*70,\computeNine*6.24) {}; \node[plotpoint] (RE2) at (axis cs:\computeNine*45,\computeNine*3.06) {}; \node[plotpoint,pin={300:REAL}] (REAL) at (axis cs:\computeNine*103,\computeNine*0.91) {}; \node[plotpoint,pin={330:ReachFlags}] (ReachFlags) at (axis cs:\computeNine*348,\computeNine*0.84) {}; \node[plotpoint,pin={[pin distance=.1cm]45:HNR}] (HNR) at (axis cs:\computeNine*15,\computeNine*0.85) {}; \node[plotpoint,pin={135:CALT}] (CALT) at (axis cs:\computeNine*11,\computeNine*1.34) {}; \node[plotpoint] at (AF) {}; \node[plotpoint] at (PHASTAF) {}; \end{scope}

\begin{scope}[-,very thick,kit-gruen!75,every node/.style={kit-schwarz,inner sep=.025cm,font=\smaller},every pin/.style={inner sep=.025cm,pin distance=.3cm,fill=white,font=\smaller}] \draw (RE) -- (RE2); \end{scope} \end{loglogaxis} \end{tikzpicture} \caption[Preprocessing and query performance of various speedup techniques]{Preprocessing and average query time performance for algorithms with available experimental data on the road network of Western Europe, using travel times as edge weights. Connecting lines indicate different trade-offs for the same algorithm. The figure is inspired by~\cite{s-spqsn-14}.} \label{fig:literature:performance} \end{figure}

Figure~\ref{fig:literature:performance} succinctly represents the performance of previously published implementations of various point-to-point algorithms on the Western Europe instance, using travel time as the cost function. For each method, the plot relates its preprocessing and average query times. Queries compute the length of the shortest path~(but not its actual list of edges) between sources and targets picked uniformly at random from the full graph. For readability, space consumption~(a third important quality measure) is not explicitly represented.\footnote{The reader is referred to Sommer~\cite{s-spqsn-14} for a similar plot~(which inspired ours) relating query times to preprocessing space.} We reproduce the numbers reported by Bauer et al.~\cite{bdsssw-chgds-10} for Reach, HH, HNR, ALT,~(bidirectional) Arc Flags, REAL, HH*, SHARC, CALT, CHASE, ReachFlags and TNR+AF. For CHASE and Arc Flags, we also consider variants with quicker PHAST-based preprocessing~\cite{dgnw-phast-13}. In addition, we consider the recent ALT implementation by Efentakis and Pfoser~\cite{ep-olbrp-13}. Moreover, we report results for several variants of TNR~\cite{bdsssw-chgds-10,als-tnrr-13}, Hub Labels~\cite{adgw-hhlsp-12,dgw-hlc-13}, HPML~\cite{dhmsw-hpmlr-09}, Contraction Hierarchies~(CH)~\cite{gssv-erlrn-12}, and Customizable Contraction Hierarchies~(CCH)~\cite{dsw-cch-sea-14}. CRP~(and the corresponding PUNCH) figures~\cite{dgpw-crprn-13} use a more realistic graph model that includes turn costs. For reference, the plot includes unidirectional and bidirectional implementations of Dijkstra's algorithm using a 4-heap. (Note that one can obtain a 20\% improvement when using a multilevel bucket queue~\cite{g-apspa-08}.) Finally, the table-lookup figure is based on the time of a single memory access in our reference machine and the precomputation time of~$\abs{\vertices}$ shortest path trees using PHAST~\cite{dgnw-phast-13}. Note that a machine with more than one petabyte of RAM~(as required by this algorithm) would likely have slower memory access times.

Times in the plot are on a single core of an Intel X5680 3.33\,GHz CPU, a mainstream server at the time of writing. Several of the algorithms in the plot were originally run on this machine~\cite{adgw-hhlsp-12,dgnw-phast-13,dgpw-crprn-13,dgw-hlc-13}; for the remaining, we divide by the following scaling factors: 2.322~for~\cite{bdsssw-chgds-10,dhmsw-hpmlr-09}, 2.698~for~\cite{gssv-erlrn-12}, 1.568~for~\cite{als-tnrr-13}, 0.837~for~\cite{dsw-cch-sea-14}, and 0.797~for~\cite{ep-olbrp-13}. These were obtained from a benchmark~(developed for this survey) that measures the time of computing several shortest path trees on the publicly available USA road network with travel times~\cite{dgj-spndi-09}. For the machines we did not have access to, we asked the authors to run the benchmark for us~\cite{ep-olbrp-13}. The benchmark is available from \url{http://algo.iti.kit.edu/~pajor/survey/}, and we encourage future works to use it as a base to compare~(sequential) running times with existing approaches.

The figure shows that there is no best technique. To stress this point, techniques with at least one implementation belonging to the Pareto set~(considering preprocessing time, query time, and space usage) are drawn as solid circles; hollow entries are dominated. The Pareto set is quite large, with various methods allowing for a wide range of space-time trade-offs. Moreover, as we shall see when examining more realistic models, these three are not the only important criteria for real-world applications.

\begin{table}[t] \centering \caption{Performance of various speedup techniques on Western Europe. Column \emph{source} indicates the implementation tested for this survey.} \label{tab:road_simple} \setlength{\tabcolsep}{.75ex} \begin{tabular}{llcrrp{.5ex}rr} \toprule &&& \multicolumn{2}{c}{\textsc{data structures}} && \multicolumn{2}{c}{\textsc{queries}}\\ \cmidrule{4-5}\cmidrule{7-8} &impl.&& space & time && scanned & time \\ algorithm &source&& [GiB] & [h:m] && vertices & [\textmu{}s] \\ \midrule Dijkstra & \cite{dgnw-phast-13} && 0.4 & -- && 9\,326\,696 & 2\,195\,080\skipDTZ\\ Bidir. Dijkstra & \cite{dgnw-phast-13} && 0.4 & -- && 4\,914\,804 & 1\,205\,660\skipDTZ\\ CRP & \cite{dgpw-crprn-13} && 0.9 & 1:00 && 2\,766 & 1\,650\skipDTZ\\ Arc Flags & \cite{dgnw-phast-13} && 0.6 & 0:20 && 2\,646 & 408\skipDTZ\\ CH & \cite{dgpw-crprn-13} && 0.4 & 0:05 && 280 & 110\skipDTZ\\ CHASE & \cite{dgnw-phast-13} && 0.6 & 0:30 && 28 & 5.76\\ HLC & \cite{dgw-hlc-13} && 1.8 & 0:50 && -- & 2.55\\ TNR & \cite{als-tnrr-13} && 2.5 & 0:22 && -- & 2.09\\ TNR+AF & \cite{bdsssw-chgds-10} && 5.4 & 1:24 && -- & 0.70\\ HL & \cite{dgw-hlc-13} && 18.8 & 0:37 && -- & 0.56\\ HL-$\infty$ & \cite{adgw-hhlsp-12} && 17.7 & 60:00 && -- & 0.25\\ table lookup & \cite{dgnw-phast-13} && 1\,208\,358.7 & 145:30 && -- & 0.06\\ \bottomrule \end{tabular} \end{table}

Table~\ref{tab:road_simple} has additional details about the methods in the Pareto set, including two versions of Dijkstra's algorithm, one Dijkstra-based hierarchical technique~(CH), three non-graph-based algorithms~(TNR, HL, HLC), and two combinations~(CHASE and TNR+AF). For reference, the table also includes a goal-directed technique~(Arc Flags) and a separator-based algorithm~(CRP), even though they are dominated by other methods. All algorithms were rerun for this survey on the reference machine~(Intel X5680 3.33\,GHz CPU), except those based on TNR, for which we report scaled results. All runs are single-threaded for this experiment, but note that all preprocessing algorithms could be accelerated using multiple cores~(and, in some cases, even GPUs)~\cite{dgnw-phast-13,gv-errnt-11}.

For each method, Table~\ref{tab:road_simple} reports the total amount of space required by all data structures~(including the graph, if needed, but excluding extra information needed for path unpacking), the total preprocessing time, the number of vertices scanned by an average query~(where applicable) and the average query time. Once again, queries consist of pairs of vertices picked uniformly at random. We note that all methods tested can be parametrized~(typically within a relatively narrow band) to achieve different trade-offs between query time, preprocessing time, and space. For simplicity, we pick a single ``reasonable'' set of parameters for each method. The only exception is HL-$\infty$, which achieves the fastest reported query times but whose preprocessing is unreasonably slow.

Observe that algorithms based on any one of the approaches considered in Section~\ref{sec:p2p} can answer queries in milliseconds or less. Separator-based~(CRP), hierarchical~(CH), and goal-directed~(Arc Flags) methods do not use much more space than Dijkstra's algorithm, but are three to four orders of magnitude faster. By combining hierarchy-based pruning and goal direction, CHASE improves query times by yet another order of magnitude, visiting little more than the shortest path itself. Finally, when a higher space overhead is acceptable, non-graph-based methods can be more than a million times faster than the baseline. In particular, HL-$\infty$ is only 5 times slower than the trivial table-lookup method, where a query consists of a single access to main memory. Note that the table-lookup method itself is impractical, since it would require more than one petabyte of RAM.

\begin{figure}[!t] \begin{tikzpicture}

\tikzset{every mark/.append style={scale=1.33}} \begin{semilogyaxis}[ width=\linewidth, height=.7\linewidth, log basis y=10, log ticks with fixed point, ymin=0.03, ymax=90000, xmin=5, xmax=25, thick, legend style={ font=\small, draw=none, fill=none, row sep=-.5ex, legend pos=north west, legend cell align=left, legend columns=2 }, grid=both, major grid style={color=kit-schwarz!15}, minor grid style={color=kit-schwarz!15,densely dotted}, xlabel={Dijkstra Rank}, ylabel={Running Time [\textmu{}s]}, minor x tick num=1, ylabel near ticks, major tick style={black, semithick}, xtick={6,8,...,24}, xticklabels={$2^6$,$2^8$,$2^{10}$,$2^{12}$,$2^{14}$,$2^{16}$,$2^{18}$,$2^{20}$,$2^{22}$,$2^{24}$}, y tick label style={rotate=90,/pgf/number format/.cd,set thousands separator={\,}}, ]

\pgfplotstableread[row sep=\\,col sep=comma]{ rank,dij,bidij,ch,chase,tnr,arcflags,hl,crp,hlinf,hlc,chtnr,lookup\\ 6,25.1823,13.8006,6.19446,4.25,19.5,9.59,0.811,25,1.088,2500.68,5.07,0.055\\ 7,32.0715,19.5812,7.58511,5.27,20,14.73,0.813,36,1.095,2494.81,6.11,0.055\\ 8,59.8574,34.5978,9.25605,6.71,20,23.92,0.817,54,1.098,2497.52,7.48,0.055\\ 9,93.8098,54.2407,11.2551,8.75,20,38.57,0.818,78,1.099,2502.18,9.15,0.055\\ 10,162.232,86.2884,13.6902,10.58,19.5,68.32,0.814,111,1.079,2486.24,10.83,0.055\\ 11,308.936,158.121,16.7633,13.92,10,123.06,0.818,149,1.092,2505.04,13.46,0.055\\ 12,602.165,290.415,20.6675,17.81,9,218.6,0.81,183,1.061,2493.01,15.65,0.055\\ 13,1170.98,549.816,25.3538,22.67,9,400.35,0.801,216,1.029,2469.25,18.26,0.055\\ 14,2373.49,1046.42,30.944,27.36,9,730.34,0.793,249,0.996,2469.25,20.71,0.055\\ 15,4802.48,2015.47,37.5752,31.32,8,1364,0.782,279,0.947,2462.03,22.05,0.055\\ 16,9991.01,3964.2,44.3654,32.12,8,2273.66,0.772,315,0.893,2454.36,19.32,0.055\\ 17,21019.8,8565.13,51.9325,21.99,6,790.06,0.752,362,0.782,2449.4,2.01,0.055\\ 18,44606.8,19481,60.203,13.79,5,558.2,0.73,419,0.629,2449.25,1.65,0.055\\ 19,94595.6,44113.4,69.2326,9.92,5,469.9,0.707,500,0.527,2434.66,1.47,0.055\\ 20,203231,107208,81.0125,6.93,5,328.64,0.674,615,0.433,2409.55,1.39,0.055\\ 21,434262,245504,91.8213,5.64,5,206.97,0.658,805,0.375,2440.98,1.37,0.055\\ 22,653116,572755,103.325,5.17,5,176.05,0.629,1167,0.308,2459.32,1.35,0.055\\ 23,923802,561073,107.144,4.94,5,175.24,0.581,1716,0.254,2457.52,1.43,0.055\\ 24,1.06E+06,6.26E+05,91.3819,4.85,5,175,0.475,2069,0.225,2191.36,1.42,0.055\\ }\dataTable

\addplot[solid,mark=*,kit-gruen] table [x=rank,y=dij] {\dataTable}; \addplot[kit-blau,mark=*,mark options={fill=white, scale=1.33}] table [x=rank,y=bidij] {\dataTable}; \addplot[solid,mark=triangle*,kit-rot] table [x=rank,y=crp] {\dataTable}; \addplot[solid,mark=diamond*,kit-maigruen] table [x=rank,y=arcflags] {\dataTable}; \addplot[solid,mark=pentagon*,kit-cyan-blau] table [x=rank,y expr=\thisrow{ch}] {\dataTable}; \addplot[solid,mark=halfcircle*,kit-orange] table [x=rank,y=chase] {\dataTable}; \addplot[solid,mark=halfsquare*,kit-braun] table [x=rank,y expr=\thisrow{hlc}/1000] {\dataTable}; \addplot[solid,mark=|,kit-lila] table [x=rank,y=chtnr] {\dataTable}; \addplot[mark=square*,kit-gruen] table [x=rank,y expr=\thisrow{hl}/1.05] {\dataTable}; \addplot[solid,mark=10-pointed star,kit-blau] table [x=rank,y expr=\thisrow{hlinf}/1.05] {\dataTable}; \addplot[solid,mark=asterisk,kit-rot] table [x=rank,y=lookup] {\dataTable};

\legend{Dijkstra,Bi.~Dijkstra,CRP,Arc Flags,CH,CHASE,HLC,TNR,HL,HL-$\infty$,T.~Lookup} \end{semilogyaxis}

\end{tikzpicture} \caption{Performance of speedup techniques for various Dijkstra ranks.} \label{fig:DijkstraRanks} \end{figure}
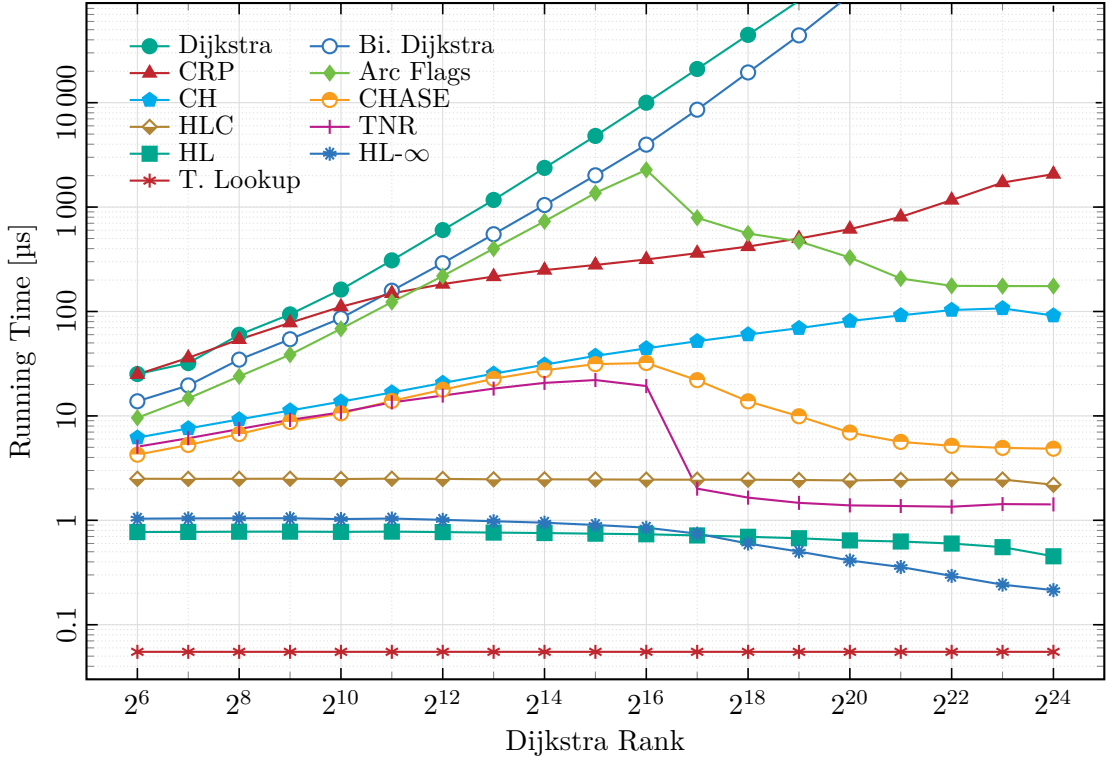

The experiments reported so far consider only random queries, which tend to be long-range. In a real system, however, most queries tend to be local. For that reason, Sanders and Schultes~\cite{ss-hhhes-05} introduced a methodology based on \emph{Dijkstra ranks}. When running Dijkstra's algorithm from a vertex~$\sourcevertex$, the rank of a vertex~$\vertex$ is the order in which it is taken from the priority queue. By evaluating pairs of vertices for Dijkstra ranks~$2^1, 2^2, \dots, 2^{\lfloor \log \abs{\vertices} \rfloor}$ for some randomly chosen sources, all types~(local, mid-range, global) of queries are evaluated. Figure~\ref{fig:DijkstraRanks} reports the median running times for all techniques from Table~\ref{tab:road_simple}~(except TNR+AF, for which such numbers have never been published) for 1\,000 random sources and Dijkstra ranks~$\geq 2^6$. As expected, algorithms based on graph searches~(including Dijkstra, CH, CRP, and Arc Flags) are faster for local queries. This is not true for bounded-hop algorithms. For TNR, in particular, local queries must actually use a~(significantly slower) graph-based approach. HL is more uniform overall because it never uses a graph.

\subsubsection{Realistic Setting} \label{ssec:practical}

Although useful, the results shown in Table~\ref{tab:road_simple} do not capture all features that are important for real-world systems. First, systems providing actual driving directions must account for turn costs and restrictions, which the simplified graph model ignores. Second, systems must often support multiple metrics~(cost functions), such as shortest distances, avoid U-turns, avoid/prefer freeways, or avoid ferries; metric-specific data structures should therefore be as small as possible. Third, query times should be robust to the choice of cost functions: the system should not time out if an unfriendly cost function is chosen. Finally, one should be able to incorporate a new cost function quickly to account for current traffic conditions~(or even user preferences).

CH has the fastest preprocessing among the algorithms in Table~\ref{tab:road_simple} and its queries are fast enough for interactive applications. Its performance degrades under realistic constraints~\cite{dgpw-crprn-13}, however. In contrast, CRP was developed with these constraints in mind. As explained in Section~\ref{sec:literature:road:separator}, it splits its preprocessing phase in two: although the initial metric-independent phase is relatively slow~(as shown in Table~\ref{tab:road_simple}), only the subsequent~(and fast) metric-dependent customization phase must be rerun to incorporate a new metric. Moreover, since CRP is based on edge separators, its performance is~(almost) independent of the cost function.

Table~\ref{tab:turns}~(reproduced from~\cite{dgpw-crprn-13}) compares CH and CRP with and without turn costs, as well as for travel distances. The instance tested is the same in Table~\ref{tab:road_simple}, augmented by turn costs~(set to 100 seconds for U-turns and zero otherwise). This simple change makes it almost as hard as fully realistic~(proprietary) map data used in production systems~\cite{dgpw-crprn-13}. The table reports metric-independent preprocessing and metric-dependent customization separately; ``DS'' refers to the data structures shared by all metrics, while ``\tabhead{custom}'' refers to the additional space and time required by each individual metric. Unlike in Table~\ref{tab:road_simple}, space consumption also includes data structures used for path unpacking. For queries, we report the time to get just the length of the shortest path~(\emph{dist}), as well as the total time to retrieve both the length and the full path~(\emph{path}). Moreover, preprocessing~(and customization) times refer to multi-threaded executions on~12~cores; queries are still sequential.

As the table shows, CRP query times are very robust to the cost function and the presence of turns. Also, a new cost function can be applied in roughly~370\,ms, fast enough to even support user-specific cost functions. Customization times can be even reduced to 36\,ms with GPUs~\cite{dkw-cddgp-14}, also reducing the amount of data stored in main memory by a factor of 6. This is fast enough for setting the cost function at \emph{query time}, enabling realistic personalized driving directions on continental scale. If GPUs are not available or space consumption is an issue, one can drop the contraction-based customization. This yields customization times of about one second on a 12-core CPU, which is still fast enough for many scenarios. In contrast, CH performance is significantly worse on metrics other than travel times without turn costs.

We stress that not all applications have the same requirements. If only good estimates on travel times~(and not actual paths) are needed, ignoring turn costs and restrictions is acceptable. In particular, ranking POIs according to travel times~(but ignoring turn costs) already gives much better results than ranking based on geographic distances. Moreover, we note that CH has fast queries even with fully realistic turn costs. If space~(for the expanded graph) is not an issue, it can still provide a viable solution to the static problem; the same holds for related methods such as HL and HLC~\cite{dgw-hlc-13}. For more dynamic scenarios, CH preprocessing can be made parallel~\cite{gv-errnt-11} or even distributed~\cite{klsv-dtdch-10}; even if run sequentially, it is fast enough for large metropolitan areas.

\begin{table}[t] \setlength{\tabcolsep}{.7ex} \small \centering \caption{Performance of Contraction Hierarchies and CRP on a more realistic instance, using different graph representations. Preprocessing and customization times are given for multi-threaded execution on a 12-core server, while queries are run single-threaded.} \label{tab:turns} \begin{tabular}{llrrrrrp{.5em}rrrrrrr} \toprule && \multitabhead{5}{CH} && \multitabhead{7}{CRP}\\ \cmidrule{3-7}\cmidrule{9-15} &&\multitabhead{2}{DS} & \multitabhead{3}{queries} && \multitabhead{2}{DS}& \multitabhead{2}{custom} & \multitabhead{3}{queries}\\ &turn& time & space & nmb. & dist & path && time & space& time & space & nmb. & dist & path\\ metric&info& [h:m] & [GiB] & scans & [ms]& [ms] && [h:m] & [GiB]& [s] & [GiB] & scans & [ms]& [ms]\\ \midrule dist&none & 0:12 & 0.68 & 858 & 0.87 & 1.07 && 0:12 & 3.11 & 0.37 & 0.07 & 2942 & 1.91 & 2.49 \\[3pt] time&none & 0:02 & 0.60 & 280 & 0.11 & 0.21 && 0:12 & 3.11 & 0.37 & 0.07 & 2766 & 1.65 & 1.81 \\ &arc-based & 0:23 & 3.14 & 404 & 0.20 & 0.30 && -- & -- & -- & -- & -- & -- & --\\ &compact & 0:29 & 1.09 &1998 & 2.27 & 2.37 && 0:12 & 3.11 & 0.37 & 0.07 & 3049 & 1.67 & 1.85 \\ \bottomrule \end{tabular} \end{table}

\subsection{Applications} \label{ssec:applications}

As discussed in Section~\ref{ssec:extensions}, many speedup techniques can handle more than plain point-to-point shortest path computations. In particular, hierarchical techniques such as CH or CRP tend to be quite versatile, with many established extensions.

Some applications may involve more than one path between a source and a target. For example, one may want to show the user several ``reasonable'' paths~(in addition to the shortest one)~\cite{camvit}. In general, these alternative paths should be short, smooth, and significantly different from the shortest path~(and other alternatives). Such paths can either be computed directly as the concatenation of partial shortest paths~\cite{camvit,ls-csarr-12,adgw-arrn-13,dgpw-crprn-13,k-hdara-13} or compactly represented as a small graph~\cite{bdgs-argrn-11,pz-iarp-13,krs-eepma-13}. A related problem is to compute a \emph{corridor}~\cite{dklw-rmrpl-12} of paths between source and target, which allows deviations from the best route~(while driving) to be handled without recomputing the entire path. These robust routes can be useful in mobile scenarios with limited connectivity. Another useful tool to reduce communication overhead in such cases is route compression~\cite{bglsz-erchr-12}.

Extensions that deal with nontrivial cost functions have also been considered. In particular, one can extend CH to handle flexible arc restrictions~\cite{grst-rpfer-12}~(such as height or weight limitations) or even multiple criteria~\cite{gks-rpfof-10,fs-pcchm-13}~(such as optimizing costs and travel time). Minimizing the energy consumption of electric vehicles~\cite{efs-orpev-11,sf-cbpvn-12,sf-eemfl-13,bdpw-eorev-13-gis,bdhpw-sctev-14,gp-tpbsf-14} is another nontrivial application, since batteries are recharged when the car is going downhill. Similarly, optimal cycling routes must take additional constraints~(such as the amount of uphill cycling) into account~\cite{s-rpbec-12}.

The ability of computing many~(batched) shortest paths fast enables interesting new applications. By quickly analyzing multiple candidate shortest paths, one can efficiently match GPS traces to road segments~\cite{efhss-ampt-11,ef-srqgt-12}. Traffic simulations also benefit from acceleration techniques~\cite{ls-hdfue-11}, since they must consider the likely routes taken by \emph{all} drivers in a network. Another application is route prediction~\cite{kh-pwdyw-07}: one can estimate where a vehicle is~(likely) headed by measuring how good its current location is as a via point towards each candidate destination. Fast routing engines allow more locations to be evaluated more frequently, leading to better predictions~\cite{adfgw-hldbl-12,efhss-ampt-11,hk-shotw-12,kgd-fdprp-13}. Planning placement of charging stations can also benefit from fast routing algorithms~\cite{fns-plsev-14}. Another important application is \emph{ride sharing}~\cite{glsnv-fdcrs-10,adfgw-hldbl-12,dl-mhrs-13}, in which one must match a ride request with the available offer in a large system, typically by minimizing drivers' detours.

Finally, batched shortest-path computations enable a wide range of point-of-interest queries~\cite{g-arptn-11,adfgw-hldbl-12,ef-srqgt-12,rt-gsagt-12,llzt-roada-12,dw-cpiqr-13,zltz-gtaei-13,EfentakisPV14}. Typical examples include finding the closest restaurant to a given location, picking the best post office to stop on the way home, or finding the best meeting point for a group of friends. Typically using the bucket-based approach~(cf.~Section~\ref{ssec:batched}), fast routing engines allow POIs to be ranked according to network-based cost functions~(such as travel time) rather than geographic distances. This is crucial for accuracy in areas with natural~(or man-made) obstacles, such as mountains, rivers, or rail tracks. Note that more elaborate POI queries must consider concatenations of shortest paths. One can handle these efficiently using an extension of the bucket-based approach that indexes pairs of vertices instead of individual ones~\cite{adfgw-hldbl-12,dw-cpiqr-13}.

\subsection{Alternative Settings}

So far, we have assumed that shortest path computations take place on a standard server with enough main memory to hold the input graph and the auxiliary data. In practice, however, it is often necessary to run~(parts of) the routing algorithm in other settings, such as mobile devices, clusters, or databases. Many of the methods we discuss can be adapted to such scenarios.

Of particular interest are mobile devices, which typically are slower and~(most importantly) have much less available RAM. This has motivated external memory implementation of various speedup techniques, such as ALT~\cite{gw-cppsp-05}, CH~\cite{ssv-mrp-08}, and time-dependent CH \cite{k-tmtdrp-13}. CH in particular is quite practical, supporting interactive queries by compressing the routing data structures and optimizing their access patterns.

Relational databases are another important setting in practice, since they allow users to formulate complex queries on the data in SQL, a popular and expressive declarative query language~\cite{ss-rbd-10}.

Unfortunately, the table-based computational model makes it hard~(and inefficient) to implement basic data structures such as graphs or even priority queues. Although some distance oracles based on geometric information could be implemented on a database~\cite{ss-qpudosn-10}, they are approximate and very expensive in terms of time and space, limiting their applicability to small instances. A better solution is to use HL, whose queries can very easily be expressed in SQL, allowing interactive applications based on shortest path computations entirely within a relational database~\cite{adfgw-hldbl-12}.

For some advanced scenarios, such as time-dependent networks, the preprocessing effort increases quite a lot compared to the time-independent scenario. One possible solution is to run the preprocessing in a distributed fashion. One can achieve an almost linear speedup as the number of machine increases, for both CH~\cite{klsv-dtdch-10} and CRP~\cite{etp-ccrsp-12}.

\section{Journey Planning in Public Transit Networks} \label{sec:public} \label{sec:literature:pt}

This section considers journey planning in~(schedule-based) public transit networks. In this scenario, the input is given by a timetable. Roughly speaking, a timetable consists of a set of stops~(such as bus stops or train platforms), a set of routes~(such as bus or train lines), and a set of trips. Trips correspond to individual vehicles that visit the stops along a certain route at a specific time of the day. Trips can be further subdivided into sequences of elementary connections, each given as a pair of~(origin/destination) stops and~(departure/arrival) times between which the vehicle travels without stopping. In addition, footpaths model walking connections (transfers) between nearby stops.

A key difference to road networks is that public transit networks are inherently \emph{time-dependent}, since certain segments of the network can only be traversed at specific, discrete points in time. As such, the first challenge concerns modeling the timetable appropriately in order to enable the computation of journeys, i.e., sequences of trips one can take within a transportation network. While in road networks computing a single shortest path~(typically the quickest journey) is often sufficient, in public transit networks it is important to solve more involved problems, often taking several optimization criteria into account. Section~\ref{sec:literature:pt:modeling} will address such modeling issues.

Accelerating queries for efficient journey planning is a long-standing problem~\cite{tulp1988trains,tulp1991searching,baumann1988buxtehude,sww-daola-00}. A large number of algorithms have been developed not only to answer basic queries fast, but also to deal with extended scenarios that incorporate delays, compute robust journeys, or optimize additional criteria, such as monetary cost.

\subsection{Modeling} \label{sec:literature:pt:modeling}

The first challenge is to model the timetable in order to enable algorithms that compute optimal journeys. Since the shortest-path problem is well understood in the literature, it seems natural to build a graph~$\graph = (\vertices, \arcs)$ from the timetable such that shortest paths in~$\graph$ correspond to optimal journeys. This section reviews the two main approaches to do so~(\emph{time-expanded} and \emph{time-dependent}), as well as the common types of problems one is interested to solve. For a more detailed overview of these topics, we refer the reader to an overview article by Müller-Hannemann et al.~\cite{mswz-tima-07}.

\paragraph{Time-Expanded Model.}

\begin{figure}[t] \begin{minipage}{\textwidth} \centering \raisebox{-.5\height}{\begin{tikzpicture} \node[arrival vertex] (aA) at (-.75,.5) {}; \node[transfer vertex] (tA) at (0,0) {}; \node[departure vertex] (dA) at (.75,0) {}; \node[transfer vertex] (tB) at (0,-1) {}; \node[departure vertex] (dB) at (.75,-1) {}; \node[transfer vertex] (tC) at (0,-2) {}; \node[departure vertex] (dC) at (.75,-2) {}; \path[stay arc] (tA) edge[<-,densely dotted] +(0,1) (tA) edge (tB) (tB) edge (tC) (tC) edge[densely dotted] +(0,-1) (aA) edge[bend left] (dA) (aA) edge[bend right] (tC); \path[connection arc] (aA) edge[<-] node {$\trip_1$} +(-1.5,.5) (dA) edge node {$\trip_1$} +(1.5,-.5) (dB) edge node {$\trip_2$} +(1.5,-.5) (dC) edge node {$\trip_3$} +(1.5,-.5); \path[departure arc] (tA) edge (dA) (tB) edge (dB) (tC) edge (dC); \end{tikzpicture} } \hfil \raisebox{-.5\height}{\begin{tikzpicture}[scale=.8] \node[stop vertex] (S1) at (60:2) {$\st_1$}; \node[stop vertex] (S2) at (180:2) {$\st_2$}; \node[stop vertex] (S3) at (300:2) {$\st_3$}; \begin{scope}[shift={(S1)}] \node[route vertex] (S1R1) at (105:1) {}; \node[route vertex] (S1R2) at (285:1) {}; \end{scope} \begin{scope}[shift={(S2)}] \node[route vertex] (S2R1) at (105:1) {}; \node[route vertex] (S2R2) at (0:1) {}; \node[route vertex] (S2R3) at (255:1) {}; \end{scope} \begin{scope}[shift={(S3)}] \node[route vertex] (S3R2) at (75:1) {}; \node[route vertex] (S3R3) at (255:1) {}; \end{scope} \path[route arc,<->] (S1R1) edge node {$\route_1$} (S2R1) (S1R2) edge node {$\route_2$} (S2R2) (S2R2) edge node {$\route_2$} (S3R2) (S2R3) edge node {$\route_3$} (S3R3) ; \foreach \s in {S1R1,S1R2,S3R2,S3R3} { \path[route arc,densely dotted,<->] (\s) edge +(1.5,0); } \foreach \s in {S2R1,S2R3} { \path[route arc,densely dotted,<->] (\s) edge +(-1.5,0); } \path[transfer arc,every node/.style={color=kit-gruen}] (S1) edge (S1R1) edge (S1R2) (S1R1) edge (S1) (S1R2) edge (S1) (S2) edge (S2R1) edge (S2R2) edge (S2R3) (S2R1) edge (S2) (S2R2) edge (S2) (S2R3) edge (S2) (S3) edge (S3R2) edge (S3R3) (S3R2) edge (S3) (S3R3) edge (S3); \end{tikzpicture} } \end{minipage} \caption{Realistic time-expanded~(left) and time-dependent~(right) models. Different vertex types are highlighted by shape: diamond~(arrival), circle~(transfer) and square~(departure) for the left figure; and circle~(stop) and square~(route) for the right figure. Connection arcs in the time-expanded model are annotated with its trips~$t_i$, and route arcs in the time-dependent model with its routes~$r_i$.} \label{fig:railwaymodels} \end{figure}
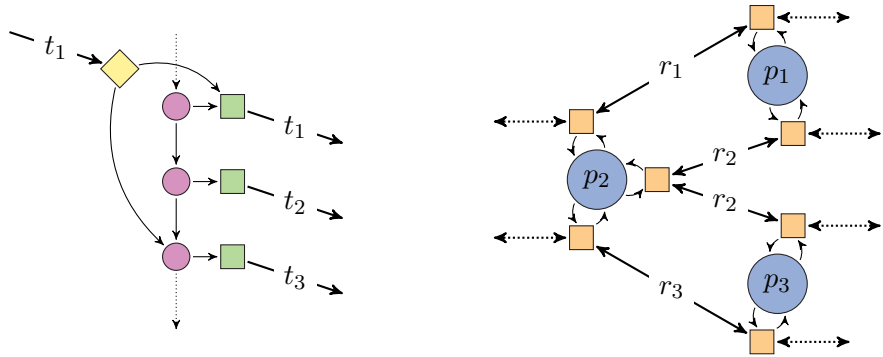

Based on the fact that a timetable consists of time-dependent \emph{events}~(e.\,g., a vehicle departing at a stop) that happen at \emph{discrete} points in time, the idea of the \emph{time-expanded} model is to build a space-time graph~(often also called an event graph)~\cite{ps-spatm-98} that ``unrolls'' time. Roughly speaking, the model creates a vertex for every event of the timetable and uses arcs to connect subsequent events in the direction of time flow. A basic version of the model~\cite{m-vvpgm-99,sww-daola-00} contains a vertex for every departure and arrival event, with consecutive departure and arrival events connected by \emph{connection} (or \emph{travel}) arcs. To enable transfers between vehicles, all vertices at the same stop are~(linearly, in chronological order) interlinked by \emph{transfer} (or \emph{waiting}) arcs. Müller-Hannemann and Weihe~\cite{mw-pspof-01} extend the model to distinguish trains~(to optimize the number of transfers taken during queries) by subdividing each connection arc by a new vertex, and then interlinking the vertices of each trip~(in order of travel). Pyrga et al.~\cite{pswz-ecspa-04,pswz-emtip-08} and Müller-Hannemann and Schnee~\cite{ms-faatc-07} extend the time-expanded model to incorporate \emph{minimum change times}~(given by the input) that are required as buffer when changing trips at a station. Their \emph{realistic} model introduces an additional \emph{transfer vertex} per departure event, and connects each arrival vertex to the first transfer vertex that obeys the minimum change time constraints. See~\figurename~\ref{fig:railwaymodels} for an illustration. If there is a footpath from stop~$\st_i$ to stop~$\st_j$, then for each arrival event at stop~$\st_i$ one adds an arc to the earliest reachable transfer vertex at~$\st_j$. This model has been further engineered~\cite{dpw-etegf-09} to reduce the number of arcs that are explored ``redundantly'' during queries.

A timetable is usually valid for a certain period of time~(up to one year). Since the timetables of different days of the year are quite similar, a space-saving technique~(\emph{compressed model}) is to consider events modulo their traffic days~\cite{msw-g-02,pswz-emtip-08}.

\paragraph{Time-Dependent Model.}

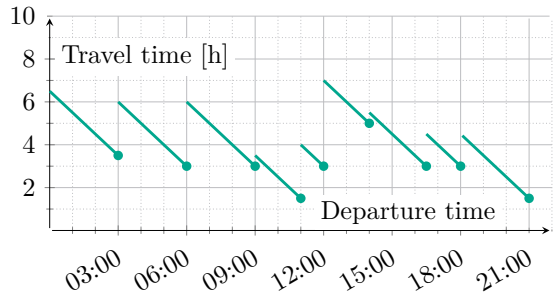
\begin{wrapfigure}{O}{0pt} \centering \begin{tikzpicture}[scale=.9] \begin{axis}[ width=9cm, height=4.75cm, axis y line=middle, axis x line=middle, date coordinates in=x, xlabel={Departure time}, ylabel={Travel time [h]}, xtick=data, xticklabel style={rotate=30,anchor=north east}, xticklabel={\hour:\minute}, xmin={1982-10-31 00:00}, xmax={1982-10-31 22:00}, ymin=0, ymax=10, date ZERO=1982-10-31, xtick={ {1982-10-31 0:00}, {1982-10-31 3:00}, {1982-10-31 6:00}, {1982-10-31 9:00}, {1982-10-31 12:00}, {1982-10-31 15:00}, {1982-10-31 18:00}, {1982-10-31 21:00}, {1982-10-31 24:00} }, minor x tick num = 2, ytick={0,2,4,6,8,10}, minor y tick num = 1, grid=both, major grid style={kit-schwarz!30}, minor grid style={densely dotted,kit-schwarz!30}, label style={fill=white,inner sep=.5ex}, ylabel style={xshift=.5ex} ] \addplot[color=kit-gruen,very thick] coordinates { (1982-10-31 00:00,6.5) (1982-10-31 03:00,3.5)

(1982-10-31 03:00,6) (1982-10-31 06:00,3)

(1982-10-31 06:00,6) (1982-10-31 09:00,3)

(1982-10-31 09:00,3.5) (1982-10-31 11:00,1.5)

(1982-10-31 11:00,4) (1982-10-31 12:00,3)

(1982-10-31 12:00,7) (1982-10-31 14:00,5)

(1982-10-31 14:00,5.5) (1982-10-31 16:30,3)

(1982-10-31 16:30,4.5) (1982-10-31 18:00,3)

(1982-10-31 18:00,4.5) (1982-10-31 21:00,1.5) }; \fill[fill=kit-gruen] (axis cs:1982-10-31 03:00,3.5) circle (2pt); \fill[fill=kit-gruen] (axis cs:1982-10-31 06:00,3) circle (2pt); \fill[fill=kit-gruen] (axis cs:1982-10-31 09:00,3) circle (2pt); \fill[fill=kit-gruen] (axis cs:1982-10-31 11:00,1.5) circle (2pt); \fill[fill=kit-gruen] (axis cs:1982-10-31 12:00,3) circle (2pt); \fill[fill=kit-gruen] (axis cs:1982-10-31 14:00,5) circle (2pt); \fill[fill=kit-gruen] (axis cs:1982-10-31 16:30,3) circle (2pt); \fill[fill=kit-gruen] (axis cs:1982-10-31 18:00,3) circle (2pt); \fill[fill=kit-gruen] (axis cs:1982-10-31 21:00,1.5) circle (2pt); \end{axis} \end{tikzpicture} \caption{Travel time function on an arc.} \label{fig:traveltimefunction} \end{wrapfigure}

The main disadvantage of the time-expanded model is that the resulting graphs are quite large~\cite{pswz-ecspa-04}. For smaller graphs, the time-dependent approach~(see Section~\ref{ssec:extensions}) has been considered by Brodal and Jacob~\cite{bj-tnmaf-04}. In their model, vertices correspond to stops, and an arc is added from~$\vertexa$ to~$\vertexb$ if there is at least one elementary connection serving the corresponding stops in this order. Precise departure and arrival times are encoded by the travel time function associated with the arc~$(\vertexa,\vertexb)$. \figurename~\ref{fig:traveltimefunction} shows the typical shape of a travel time function: each filled circle represents an elementary connection; the line segments~(with slope~$-1$) reflect not only the travel time, but also the waiting time until the next departure. Pyrga et al.~\cite{pswz-emtip-08} further extended this basic model to enable minimum change times by creating, for each stop~$\st$ and each route that serves~$\st$, a dedicated \emph{route vertex}. Route vertices at~$\st$ are connected to a common \emph{stop vertex} by arcs with constant cost depicting the minimum change time of~$\st$. Trips are distributed among \emph{route arcs} that connect the subsequent route vertices of a route, as shown in~\figurename~\ref{fig:railwaymodels}. They also consider a model that allows arbitrary minimum change times between pairs of routes within each stop~\cite{pswz-emtip-08}. Footpaths connecting nearby stops are naturally integrated into the time-dependent model~\cite{dms-mcspt-08}. For some applications, one may merge route vertices of the same stop as long as they never connect trips such that a transfer between them violates the minimum change time~\cite{dkp-pcbcp-12}.

\paragraph{Frequency-Based Model.} In real-world timetables trips often operate according to specific frequencies at times of the day. For instance, a bus may run every 5 minutes during rush hour, and every 10 minutes otherwise. Bast and Storandt~\cite{bs-fbspt-14} exploit this fact in the \emph{frequency-based model}: as in the time-dependent approach, vertices correspond to stops, and an arc between a pair of stops~$(u,v)$ is added if there is at least one elementary connection from~$u$ to~$v$. However, instead of storing the departures of an arc explicitly, those with coinciding travel times are compressed into a set of tuples consisting of an initial departure time~$\deptime$, a time interval~$\timespan$, and a frequency~$\frequency$. The corresponding original departures can thus be reconstructed by computing each~$\deptime + \frequency i$ for those~$i \in \naturals$ that satisfy~$\deptime + \frequency i\leq \deptime+\timespan$. Bast and Storandt compute these tuples by covering the set of departure times by a small set of overlapping arithmetic progressions, then discarding duplicate entries~(occurring after decompression) at query time~\cite{bs-fbspt-14}.

\paragraph{Problem Variants.} Most research on road networks has focused on computing the shortest path according to a given cost function~(typically travel times). For public transit networks, in contrast, there is a variety of natural problem formulations.

The simplest variant is the \emph{earliest arrival problem}. Given a source stop~$\sourcestop$, a target stop~$\targetstop$, and a departure time~$\sourcetime$, it asks for a journey that departs~$\sourcestop$ no earlier than~$\sourcetime$ and arrives at~$\targetstop$ as early as possible. A related variant is the \emph{range}~(or \emph{profile}) \emph{problem}~\cite{n-t-95}, which replaces the departure time by a time range~(e.\,g.~8--10\,am, or the whole day). This problem asks for a set of journeys of minimum travel time that depart within that range.

Both the earliest arrival and the range problems only consider~(arrival or travel) time as optimization criterion. In public-transit networks, however, other criteria~(such as the number of transfers) are just as important, which leads to the \emph{multicriteria problem}~\cite{mw-pspof-01}. Given source and target stops~$\sourcestop,\targetstop$ and a departure time~$\sourcetime$ as input, it asks for a~(maximal) Pareto set~$\journeys$ of nondominating journeys with respect to the optimization criteria considered. A journey~$\journey_1$ is said to dominate journey~$\journey_2$ if~$\journey_1$ is better than or equal to~$\journey_2$ in all criteria. Further variants of the problem relax or strengthen these domination rules~\cite{ms-faatc-07}.

\subsection{Algorithms without Preprocessing}

This section discusses algorithms that can answer queries without a preprocessing phase, which makes them a good fit for dynamic scenarios that include delays, route changes, or train cancellations. We group the algorithms by the problems they are meant to solve.

\paragraph{Earliest Arrival Problem.}

Earliest arrival queries on the time-expanded model can be answered in a straightforward way by Dijkstra's algorithm~\cite{sww-daola-00}, in short~TED~(time-expanded Dijkstra). It is initialized with the vertex that corresponds to the earliest event of the source stop~$\sourcestop$ that occurs after~$\sourcetime$~(in the realistic model, a transfer vertex must be selected). The first scanned vertex associated with the target stop~$\targetstop$ then represents the earliest arrival~$\sourcevertex$--$\targetvertex$ journey. In the compressed time-expanded model, slight modifications to Dijkstra's algorithm are necessary because an event vertex may appear several times on the optimal shortest path (namely for different consecutive days). One possible solution is to use a bag of labels for each vertex as in the multicriteria variants described below. Another solution is described in Pyrga et al.~\cite{pswz-emtip-08}.

On time-dependent graphs, Dijkstra's algorithm can be augmented to compute shortest paths~\cite{ch-tsrtn-66,d-aassp-69}, as long as the cost functions are nonnegative and FIFO~\cite{or-spmda-90,or-mwptd-91}. The only modification is that, when the algorithm scans an arc~$(\vertexa,\vertexb)$, the arc cost is evaluated at time~$\sourcetime + \dist(\sourcevertex,\vertexa)$. Note that the algorithm retains the label-setting property, i.\,e., each vertex is scanned at most once. In the time-dependent public transit model, the query is run from the stop vertex corresponding to~$\sourcestop$ and the algorithm may stop as soon as it extracts~$\targetstop$ from the priority queue. The algorithm is called~TDD~(time-dependent Dijkstra).

Another approach is to exploit the fact that the time-expanded graph is directed and acyclic. (Note that overnight connections can be handled by unrolling the timetable for several consecutive periods.) By scanning vertices in topological order, arbitrary queries can be answered in linear time. This simple and well-known observation has been applied for journey planning by Mellouli and Suhl~\cite{MellouliSuhl2006}, for example. While this idea saves the relatively expensive priority queue operations of Dijkstra's algorithm, one can do even better by not maintaining the graph structure explicitly, thus improving locality and cache efficiency. The recently developed \emph{Connection Scan Algorithm}~(CSA)~\cite{dpsw-isftr-13} organizes the elementary connections of the timetable in a single array, sorted by departure time. The query then only scans this array once, which is very efficient in practice. Note that CSA requires footpaths in the input to be closed under transitivity to ensure correctness.

\paragraph{Range Problem.}

The range problem can be solved on the time-dependent model by variants of Dijkstra's algorithm. The first variant~\cite{n-t-95,d-ctdsp-99} maintains, at each vertex~$\vertex$, a travel-time function~(instead of a scalar label) representing the optimal travel times from~$\sourcevertex$ to~$\vertex$ for the considered time range. Whenever the algorithm relaxes an arc~$(\vertexa,\vertexb)$, it first \emph{links} the full travel-time function associated with~$\vertexa$ to the~(time-dependent) cost function of the arc~$(\vertexa,\vertexb)$, resulting in a function that represents the times to travel from~$\sourcevertex$ to~$\vertexb$ via~$\vertexa$. This function is then \emph{merged} into the~(tentative) travel time function associated with~$\vertexb$, which corresponds to taking the element-wise minimum of the two functions. The algorithm loses the label-setting property, since travel time functions cannot be totally ordered. As a result the algorithm may reinsert vertices into the priority queue whenever it finds a journey that improves the travel time function of an already scanned vertex.

Another algorithm~\cite{b-mpq-12} exploits the fact that trips depart at discrete points in time, which helps to avoid redundant work when propagating travel time functions. When it relaxes an arc, it does not consider the full function, but each of its encoded connections individually. It then only propagates the parts of the function that have improved.

The \emph{Self-Pruning Connection Setting} algorithm~(SPCS)~\cite{dkp-pcbcp-12} is based on the observation that \emph{any} optimal journey from~$\sourcevertex$ to~$\targetvertex$ has to start with one of the trips departing from~$\sourcevertex$. It therefore runs, for each such trip, Dijkstra's algorithm from~$\sourcevertex$ at its respective departure time. SPCS performs these runs simultaneously using a shared priority queue whose entries are ordered by arrival time. Whenever the algorithm scans a vertex~$\vertex$, it checks if~$\vertex$ has been already scanned for an associated~(departing) trip with a \emph{later} departure time~(at~$\sourcevertex$), in which case it prunes~$\vertex$. Moreover, SPCS can be parallelized by assigning different subsets of departing trips from~$\sourcevertex$ to different CPU cores.

Bast and Storandt~\cite{bs-fbspt-14} propose an extension of Dijkstra's algorithm that operates on the~(compressed) frequency-based model directly. It maintains with every vertex~$\vertex$ a set of tuples consisting of a time interval, a frequency, and the travel time. Hence, a single tuple may represent multiple optimal journeys, each departing within the tuple's time interval. Whenever the algorithm relaxes an arc~$(\vertexa,\vertexb)$, it first extends the tuples from the bag at~$u$ with the ones stored at the arc~$(\vertexa,\vertexb)$ in the compressed graph. The resulting tentative bag of tuples~(representing all optimal journeys to~$v$ via~$u$) is then \emph{merged} into the bag of tuples associated with~$\vertexb$. The main challenge of this algorithm is efficiently merging tuples with incompatible frequencies and time intervals~\cite{bs-fbspt-14}.

Finally, the Connection Scan Algorithm has been extended to the range problem~\cite{dpsw-isftr-13}. It uses the same array of connections, ordered by departure time, as for earliest arrival queries. It still suffices to scan this array once, even to obtain optimal journeys to all stops of the network.

\paragraph{Multicriteria Problem.}

Although Pareto sets can contain exponentially many solutions~(see Section~\ref{ssec:extensions}), they are often much smaller for public transit route planning, since common optimization criteria are positively correlated. For example, for the case of optimizing earliest arrival time and number of transfers, the \emph{Layered Dijkstra}~(LD) algorithm~\cite{bj-tnmaf-04,pswz-emtip-08} is efficient. Given an upper bound~$\maxtransfers$ on the number of transfers, it~(implicitly) copies the timetable graph into~$\maxtransfers$ layers, rewiring transfer arcs to point to the next higher level. It then suffices to run a time-dependent~(single criterion) Dijkstra query from the lowest level to obtain Pareto sets.

In the time-expanded model, Müller-Hannemann and Schnee~\cite{ms-faatc-07} consider the Multicriteria Label-Setting~(MLS) algorithm~(cf.~Section~\ref{ssec:extensions}) to optimize arrival time, ticket cost, and number of transfers. In the time-dependent model, Pyrga et al.~\cite{pswz-emtip-08} compute Pareto sets of journeys for arrival time and number of transfers. Disser et al.~\cite{dms-mcspt-08} propose three optimizations to MLS that reduce the number of queue operations: hopping reduction, label forwarding, and dominance by early results~(or \emph{target pruning}). Bast and Storandt~\cite{bs-fbspt-14} extend the frequency-based range query algorithm to also include number of transfers as criterion.

A different approach is \emph{RAPTOR}~(Round-bAsed Public Transit Optimized Router)~\cite{dpw-rbptr-12}. It is explicitly developed for public transit networks and its basic version optimizes arrival time and the number of transfers taken. Instead of using a graph, it organizes the input as a few simple arrays of trips and routes. Essentially, RAPTOR is a dynamic program: it works in rounds, with round~$i$ computing earliest arrival times for journeys that consist of exactly~$i$ transfers. Each round takes as input the stops whose arrival time improved in the previous round~(for the first round this is only the source stop). It then \emph{scans} the routes served by these stops. To scan route~$\route$, RAPTOR traverses its stops in order of travel, keeping track of the earliest possible trip~(of~$\route$) that can be taken. This trip may improve the tentative arrival times at subsequent stops of route~$\route$. Note that RAPTOR scans each route at most once per round, which is very efficient in practice~(even faster than Dijkstra's algorithm with a single criterion). Moreover, RAPTOR can be parallelized by distributing non-conflicting routes to different CPU cores. It can also be extended to handle range queries~(rRAPTOR) and additional optimization criteria~(McRAPTOR). Note that, like CSA, RAPTOR also requires footpaths in the input to be closed under transitivity.

\subsection{Speedup Techniques}\label{sec:transit-speedup}

This section presents an overview of preprocessing-based speedup techniques for journey planning in public transit networks. A natural~(and popular) approach is to adapt methods that are effective on road networks~(see~\figurename~\ref{fig:literature:performance}). Unfortunately, the speedups observed in public transit networks are several orders of magnitude lower than in road networks. This is to some extent explained by the quite different structural properties of public transit and road networks~\cite{b-cpttw-09}. For example, the neighborhood of a stop can be much larger than the number of road segments incident to an intersection. Even more important is the effect of the inherent time-dependency of public transit networks. Thus, developing efficient preprocessing-based methods for public transit remains a challenge.

Some road network methods were tested on public transit graphs without performing realistic queries~(i.\,e., according to one of the problems from Section~\ref{sec:literature:pt:modeling}). Instead, such studies simply perform point-to-point queries on public-transit graphs. In particular, Holzer et al.~\cite{hsww-cstsp-06} evaluate basic combinations of bidirectional search, goal directed search, and Geometric Containers on a simple stop graph~(with average travel times). Bauer et al.~\cite{bdw-essut-11} also evaluated bidirectional search, ALT, Arc Flags, Reach, REAL, Highway Hierarchies, and SHARC on time-expanded graphs. Core-ALT, CHASE, and Contraction Hierarchies have also been evaluated on time-expanded graphs~\cite{bdsssw-chgds-10}.

\paragraph{\astar Search.} On public transit networks, basic \astar~search has been applied to the time-dependent model~\cite{pswz-emtip-08,dms-mcspt-08}. In the context of multicriteria optimization, Disser et al.~\cite{dms-mcspt-08} determine lower bounds for each vertex~$\vertex$~to the target stop~$\targetstop$~(before the query) by running a backward search~(from~$\targetstop$) using~the~(constant) lower bounds of the travel time functions as arc cost.

\paragraph{ALT.} The~(unidirectional) ALT~\cite{gh-cspas-05} algorithm has been adapted to both the time-expanded~\cite{dpw-etegf-09} and the time-dependent~\cite{ndls-bastd-12} models for computing earliest arrival queries. In both cases, landmark selection and distance precomputation is performed on an auxiliary stop graph, in which vertices correspond to stops and an arc is added between two stops~$\st_i,\st_j$ if there is an elementary connection from~$\st_i$ to~$\st_j$ in the input. Arc costs are lower bounds on the travel time between their endpoints.

\paragraph{Geometric Containers.} Geometric containers~\cite{sww-daola-00,wwz-gcesp-05} have been extensively tested on the time-expanded model for computing earliest arrival queries. In fact, they were developed in the context of this model. As mentioned in Section~\ref{sec:p2p}, bounding boxes perform best~\cite{wwz-gcesp-05}.

\paragraph{Arc Flags and SHARC.} Delling et al.~\cite{dpw-etegf-09} have adapted Arc Flags~\cite{l-aeept-09,hkms-fppsp-09} to the time-expanded model as follows. First, they compute a partition on the stop graph~(defined as in ALT). Then, for each boundary stop~$\st$ of cell~$\cell$, and each of its arrival vertices, a backward search is performed on the time-expanded graph. The authors observe that public transit networks have many paths of equal length between the same pair of vertices~\cite{dpw-etegf-09}, making the choice of tie-breaking rules important. Furthermore, Delling et al.~\cite{dpw-etegf-09} combine Arc Flags, ALT, and a technique called \emph{Node Blocking}, which avoids exploring multiple arcs from the same route.

SHARC, which combines Arc Flags with shortcuts~\cite{bd-sharc-09}, has been tested on the time-dependent model with earliest arrival queries by Delling~\cite{d-tdsr-11}. Moreover, Arc Flags with shortcuts for the Multi-Label-Setting algorithm~(MLS) have been considered for computing full~(i.\,e., using strict domination) Pareto sets using arrival time and number of transfers as criteria~\cite{bdgm-atdmc-09}. In time-dependent graphs, a flag must be set if its arc appears on a shortest path toward the corresponding cell at least once during the time horizon~\cite{d-tdsr-11}. For better performance, one can use different sets of flags for different time periods~(e.\,g., every two hours). The resulting total speedup is still below 15, from which it is concluded that ``accelerating time-dependent multicriteria timetable information is harder than expected''~\cite{bdgm-atdmc-09}. Slight additional speedups can be obtained if one restricts the search space to only those solutions in the Pareto set for which the travel time is within an interval defined by the earliest arrival time and some upper bound. Berger et al.~\cite{bgm-fdsut-10} observed that in such a scenario optimal substructure in combination with lower travel time bounds can be exploited and yield additional pruning during search. It is worth noting that this method does not require any preprocessing and is therefore well-suited for a dynamic scenario.

\paragraph{Overlay Graphs.} To accelerate earliest arrival queries, Schulz et al.~\cite{sww-daola-00} compute single-level overlays between ``important'' hub stations in the time-expanded model, with importance values given as input. More precisely, given a subset of important stations, the overlay graph consists of \emph{all} vertices~(events) that are associated with these stations. Edges in the overlay are computed such that distances between any pair of vertices~(events) are preserved. Extending this approach to overlay graphs over multiple levels of hub stations~(selected by importance or degree) results in speedups of about 11~\cite{swz-umlgt-02}.

\paragraph{Separator-based techniques.} Strasser and Wagner~\cite{sw-csa-13} combine the Connection Scan Algorithm~\cite{dpsw-isftr-13} with ideas of customizable route planning~(CRP)~\cite{dgpw-crprn-13} resulting in the Accelerated Connection Scan Algorithm~(ACSA). It is designed for both earliest arrival and range queries. ACSA first computes a multilevel partition of stops, minimizing the number of elementary connections with endpoints in different cells. Then, it precomputes for each cell the partial journeys~(transit connections) that cross the respective cell. For queries, the algorithm essentially runs CSA restricted to the elementary connections of the cells containing the source or target stops, as well as transit connections of other~(higher-level) cells. As shown in Section~\ref{sec:pt:experiments}, it achieves excellent query and preprocessing times on country-sized instances.

\paragraph{Contraction Hierarchies.} The Contraction Hierarchies algorithm~\cite{gssv-erlrn-12} has been adapted to the realistic time-dependent model with minimum change times for computing earliest arrival and range queries~\cite{g-ctnrt-10}. It turns out that simply applying the algorithm to the route model graph results in too many shortcuts to be practical. Therefore, contraction is performed on a condensed graph that contains only a single vertex per stop. Minimum change times are then ensured by the query algorithm, which must maintain multiple labels per vertex.

\paragraph{Transfer Patterns.} A speedup technique specifically developed for public transit networks is called \emph{Transfer Patterns}~\cite{bceghrv-frvlp-10}. It is based on the observation that many optimal journeys share the same transfer pattern, defined as the sequence of stops where a transfer occurs. Conceptually, these transfer patterns are precomputed using range queries for all pairs of stops and departure times. At query time, a query graph is built as the union of the transfer patterns between the source and target stops. The arcs in the query graph represent direct connections between stops~(without transfers), and can be evaluated very fast. Dijkstra's algorithm~(or MLS) is then applied to this much smaller query graph.

If precomputing transfer patterns between \emph{all} pairs of stops is too expensive, one may resort to the following two-level approach. It first selects a subset of~(important) hub stops. From the hubs, global transfer patterns are precomputed to all other stops. For the non-hubs, local transfer patterns are computed only towards relevant hub stops. This approach is similar to TNR, but the idea is applied asymmetrically: transfer patterns are computed from all stops to the hub stops, and from the hub stops to everywhere. If preprocessing is still impractical, one can restrict the local transfer patterns to at most three legs~(two transfers). Although this restriction is heuristic, the algorithm still almost always finds the optimal solution in practice, since journeys requiring more than two transfers to reach a hub station are rare~\cite{bceghrv-frvlp-10}.

\paragraph{TRANSIT.} Finally, Transit Node Routing~\cite{bfss-frrnt-07,bfm-uspqt-09,ss-racts-09} has been adapted to public transit journey planning in~\cite{aw-fmopm-12}. Preprocessing of the resulting \emph{TRANSIT} algorithm uses the~(small) stop graph to determine a set of transit nodes~(with a similar method as in~\cite{bfm-uspqt-09}), between which it maintains a distance table that contains sets of journeys with minimal travel time~(over the day). Each stop~$\st$ maintains, in addition, a set of access nodes~$\accessnodes(\st)$, which is computed on the time-expanded graph by running local searches from each departure event of~$\st$ toward the transit stops. The query then uses the access nodes of~$\sourcestop$ and~$\targetstop$ and the distance table to resolve global requests. For local requests, it runs goal-directed~\astar search. Queries are slower than for Transfer Patterns.

\subsection{Extended Scenarios}\label{sec:transit-extended}

Besides computing journeys according to one of the problems from Section~\ref{sec:literature:pt:modeling}, extended scenarios~(such as incorporating delays) have been studied as well.

\paragraph{Uncertainty and Delays.} Trains, buses and other means of transport are often prone to delays in the real world. Thus, handling delays~(and other sources of uncertainty) is an important aspect of a practical journey planning system. Firmani et al.~\cite{fils-itrar-13} recently presented a case study for the public transport network of the metropolitan area of Rome. They provide strong evidence that computing journeys according to the published timetable often fails to deliver optimal or even high-quality solutions. However, incorporating real-time GPS location data of vehicles into the journey planning algorithm helps improve the journey quality~(e.\,g., in terms of the experienced delay)~\cite{ais-egpsd-14,dips-btrev-14}.

Müller-Hannemann and Schnee~\cite{ms-etipd-09} consider the online problem where delays, train cancellations, and extra trains arrive as a continuous stream of information. They present an approach which quickly updates the time-expanded model to enable queries according to current conditions. Delling et al.~\cite{dgwz-tiucd-08} also discuss updating the time-dependent model and compare the required effort with the time-expanded model. Cionini et al.~\cite{cddfgp-egbmd-14} propose a new graph-based model which is tailored to handle dynamic updates, and they experimentally show its effectiveness in terms of both query and update times. Berger et al.~\cite{bgmo-sdplt-11} propose a realistic stochastic model that predicts how delays propagate through the network. In particular, this model is evaluated using real~(delay) data from Deutsche Bahn. Bast et al.~\cite{bss-drtpp-13} study the robustness of Transfer Patterns with respect to delays. They show that the transfer patterns computed for a scenario without any delays give optimal results for 99\,\% of queries, even when large and area-wide~(random) delays are injected into the networks.

Disser et al.~\cite{dms-mcspt-08} and Delling et al.~\cite{raptorjournal} study the computation of \emph{reliable} journeys via multicriteria optimization. The reliability of a transfer is defined as a function of the available buffer time for the transfer. Roughly speaking, the larger the buffer time, the more likely it is that the transfer will be successful. According to this notion, transfers with a high chance of success are still considered reliable even if there is no backup alternative in case they fail.

To address this issue, Dibbelt et al.~\cite{dpsw-isftr-13} minimize the \emph{expected arrival time}~(with respect to a simple model for the probability that a transfer breaks). Instead of journeys, their method~(which is based on the CSA algorithm) outputs a \emph{decision graph} representing optimal instructions to the user at each point of their journey, including cases in which a connecting trip is missed. Interestingly, minimizing the expected arrival time implicitly helps minimizing the number of transfers, since each ``unnecessary'' transfer introduces additional uncertainty, hurting the expected arrival time.

Finally, Goerigk et al.~\cite{gkmss-tpslr-13} study the computation of \emph{robust} journeys, considering both strict robustness~(i.\,e., computing journeys that are always feasible for a given set of delay scenarios) and light robustness~(i.\,e., computing journeys that are most reliable when given some extra slack time). While strict robustness turns out to be too conservative in practice, the notion of light robustness seems more promising. \emph{Recoverable robust} journeys~(which can always be updated when delays occur) have recently been considered in~\cite{ghms-rrti-13}. A different, new robustness concept has been proposed by B\"ohmov{\'a} et al.~\cite{bmpvw-rrupt-13}. In order to propose solutions that are robust for typical delays, past observations of real traffic situations are used. Roughly speaking, a route is more robust the better it has performed in the past under different scenarios.

\paragraph{Night Trains.} Gunkel et al.~\cite{gsm-h-11} have considered the computation of overnight train journeys, whose optimization goals are quite different from regular ``daytime'' journeys. From a customer's point of view, the primary objective is usually to have a reasonably long sleeping period. Moreover, arriving too early in the morning at the destination is often not desired. Gunkel et al.~present two approaches to compute overnight journeys. The first approach explicitly enumerates all overnight trains~(which are given by the input) and computes, for each such train, the optimal feeding connections. The second approach runs multicriteria search with sleeping time as a maximization criterion.

\paragraph{Fares.} Müller-Hannemann and Schnee~\cite{ms-pltcm-06} have analyzed several pricing schemes, integrating them as an optimization criterion~(cost) into MOTIS, a multicriteria search algorithm that works on the time-expanded model. In general, however, optimizing exact monetary cost is a challenging problem, since real-world pricing schemes are hard to capture by a mathematical model~\cite{ms-pltcm-06}.

Delling et al.~\cite{dpw-rbptr-12} consider computing Pareto sets of journeys that optimize fare zones with the McRAPTOR algorithm. Instead of using~(monetary) cost as an optimization criterion directly, they compute all nondominated journeys that traverse different combinations of fare zones, which can then be evaluated by cost in a quick postprocessing step.

\paragraph{Guidebook Routing.} Bast and Storandt~\cite{bs-fbgr-14} introduce \emph{Guidebook Routing}, where the user specifies only source and target stops, but neither a day nor a time of departure. The desired answer is then a set of routes, each of which is given by a sequence of train or bus numbers and transfer stations. For example, an answer may read like \emph{take bus number 11 towards the bus stop at X, then change to bus number 13 or 14~(whichever comes first) and continue to the bus stop at Y}. Guidebook routes can be computed by first running a multicriteria range query, and then extracting from the union of all Pareto-optimal time-dependent paths a subset of routes composed by arcs which are most frequently used. The Transfer Patterns algorithm lends itself particularly well to the computation of such guidebook routes. For practical guidebook routes~(excluding ``exotic'' connections at particular times), the preprocessing space and query times of Transfer Patterns can be reduced by a factor of 4 to 5.

\subsection{Experiments and Comparison} \label{sec:pt:experiments}

This section compares the performance of some of the journey planning algorithms discussed in this section. As in road networks, all algorithms have been carefully implemented in C++ using mostly custom-built data structures.

Table~\ref{tab:public_transit} summarizes the results. Running times are obtained from a sequential execution on one core of a dual 8-core Intel Xeon E5-2670 machine clocked at 2.6\;GHz with 64\,GiB of DDR3-1600 RAM. The exceptions are Transfer Patterns and Contraction Hierarchies, for which we reproduce the values reported in the original publication~(obtained on a comparable machine).

For each algorithm, we report the instance on which it has been evaluated, as well as its total number of elementary connections~(a proxy for size) and the number of consecutive days covered by the connections. Unfortunately, realistic benchmark data of country scale~(or larger) has not been widely available to the research community. Some metropolitan transit agencies have recently started making their timetable data publicly available, mostly using the General Transit Feed format\footnote{\url{https://developers.google.com/transit/gtfs/}}. Still, research groups often interpret the data differently, making it hard to compare the performance of different algorithms. The largest metropolitan instance currently available is the full transit network of London\footnote{\url{http://data.london.gov.uk/}}. It contains approximately 21 thousand stops, 2.2 thousand routes, 133 thousand trips, 46 thousand footpaths, and 5.1 million elementary connections for one full day. We therefore use this instance for the evaluation of most algorithms. The instances representing Germany and long-distance trains in Europe are generated in a similar way, but from proprietary data.

\begin{table}[!t] \centering \newcommand{\notreported}{n/a} \newcommand{\notapplicable}{--} \caption{Performance of various public transit algorithms on random queries. For each algorithm, the table indicates the implementation tested~(which may not be the publication introducing the algorithm), the instance it was tested on, its total number of elementary connections~(in millions) as well as the number of consecutive days they cover. A ``p'' indicates that the timetable is periodic (with a period of one day). The table then shows the criteria that are optimized~(a subset of arrival times, transfers, full range, fares, and reliability), followed by total preprocessing time, average number of comparisons per stop, average number of journeys in the Pareto set, and average query times in milliseconds. Missing entries either do not apply~(--) or are well-defined but not available~(\notreported). } \label{tab:public_transit} \newcommand{\noprep}{--} \newcommand{\exactalgo}{--} \newcommand{\instLondon}{London} \newcommand{\instEuropeLongDist}{Europe\,(lng)} \newcommand{\instMadrid}{Madrid} \newcommand{\instGermany}{Germany} \newcommand{\instNY}{New York} \newcommand{\instNA}{N.\,America} \newcommand{\rotation}{70} \newcommand{\rotated}[1]{\begin{rotate}{\rotation}\hspace{-0.2em}#1\end{rotate}} \newcommand{\showsource}[1]{#1} \setlength{\tabcolsep}{.66ex} \begin{tabular}{lrlrrccccccrrrr} \toprule && \multicolumn{3}{c}{\textsc{input}} && \multicolumn{5}{c}{\textsc{criteria}} & & \multicolumn{3}{c}{\textsc{query}} \\ \cmidrule{3-5} \cmidrule{7-11} \cmidrule{13-15} && & conn. & && & & & & & prep. & comp. & & time \\ algorithm &\multicolumn{1}{c}{\rotated{impl.}}& name & [$10^6$] & dy. && \rotated{arr.} & \rotated{tran.} & \rotated{rng.} & \rotated{fare} & \rotated{rel.} & [h] & /stop & jn. & [ms]\\ \midrule TED & & \instLondon & 5.1 & 1 && \enabled & \disabled & \disabled & \disabled & \disabled & \noprep & 50.6 & 0.9 & 44.8 \\ TDD&\showsource{~\cite{raptorjournal}} & \instLondon & 5.1 & 1 && \enabled & \disabled & \disabled & \disabled & \disabled & \noprep & 7.4 & 0.9 & 11.0 \\ CH&\showsource{~\cite{g-ctnrt-10}} & \instEuropeLongDist & 1.7 & p && \enabled & \disabled & \disabled & \disabled & \disabled & $<$\,0.1 & $<$\,0.1 & \notreported & 0.3\\ CSA&\showsource{~\cite{dpsw-isftr-13}} & \instLondon & 4.9 & 1 && \enabled & \disabled & \disabled & \disabled & \disabled & \noprep & 26.6 & \notreported & 2.0\\ ACSA&\showsource{~\cite{sw-csa-13}} & \instGermany & 46.2 & 2 && \enabled & \disabled & \disabled & \disabled & \disabled & \noprep & \notreported & \notreported & 8.7\\ T.\,Patterns&\showsource{~\cite{bs-fbspt-14}} & \instGermany & 90.4 & 7 && \enabled & \disabled & \disabled & \disabled & \disabled & 541 & \notapplicable & 1.0 & 0.4 \\ \addlinespace LD&\showsource{~\cite{raptorjournal}} & \instLondon & 5.1 & 1 && \enabled & \enabled & \disabled & \disabled & \disabled & \noprep & 15.6 & 1.8 & 28.7\\ MLS&\showsource{~\cite{raptorjournal}} & \instLondon & 5.1 & 1 && \enabled & \enabled & \disabled & \disabled & \disabled & \noprep & 23.7 & 1.8 & 50.0\\ RAPTOR&\showsource{~\cite{raptorjournal}} & \instLondon & 5.1 & 1 && \enabled & \enabled & \disabled & \disabled & \disabled & \noprep & 10.9 & 1.8 & 5.4\\ T.\,Patterns&\showsource{~\cite{bs-fbspt-14}} & \instGermany & 90.4 & 7 && \enabled & \enabled & \disabled & \disabled & \disabled & 566 & \notapplicable & 2.0 & 0.8 \\ \addlinespace CH&\showsource{~\cite{g-ctnrt-10}} & \instEuropeLongDist & 1.7 & p && \enabled & \disabled & \enabled & \disabled & \disabled & $<$\,0.1 & $<$\,0.1 & \notreported & 3.7 \\ SPCS&\showsource{~\cite{dpsw-isftr-13}} & \instLondon & 4.9 & 1 && \enabled & \disabled & \enabled & \disabled & \disabled & \noprep & 372.5 & 98.2 & 843.0 \\ CSA&\showsource{~\cite{dpsw-isftr-13}} & \instLondon & 4.9 & 1 && \enabled & \disabled & \enabled & \disabled & \disabled & \noprep & 436.9 & 98.2 & 161.0 \\ ACSA&\showsource{~\cite{sw-csa-13}} & \instGermany & 46.2 & 2 && \enabled & \disabled & \enabled & \disabled & \disabled & 8 & \notreported & \notreported & 171.0\\ T.\,Patterns&\showsource{~\cite{bs-fbspt-14}} & \instGermany & 90.4 & 7 && \enabled & \disabled & \enabled & \disabled & \disabled & 541 & \notapplicable & 121.2 & 22.0 \\ \addlinespace rRAPTOR&\showsource{~\cite{dpsw-isftr-13}} & \instLondon & 4.9 & 1 && \enabled & \enabled & \enabled & \disabled & \disabled & \noprep & 1634.0 & 203.4 & 922.0\\ CSA&\showsource{~\cite{dpsw-isftr-13}} & \instLondon & 4.9 & 1 && \enabled & \enabled & \enabled & \disabled & \disabled & \noprep & 3824.9 & 203.4 & 466.0 \\ T.\,Patterns&\showsource{~\cite{bs-fbspt-14}} & \instGermany & 90.4 & 7 && \enabled & \enabled & \enabled & \disabled & \disabled & 566 & \notapplicable & 226.0 & 39.6 \\ \addlinespace MLS&\showsource{~\cite{raptorjournal}} & \instLondon & 5.1 & 1 && \enabled & \enabled & \disabled & \enabled & \disabled & \noprep & 818.2 & 8.8 & 304.2 \\ McRAPTOR&\showsource{~\cite{raptorjournal}} & \instLondon & 5.1 & 1 && \enabled & \enabled & \disabled & \enabled & \disabled & \noprep & 277.5 & 8.8 & 100.9 \\ \addlinespace MLS&\showsource{~\cite{raptorjournal}} & \instLondon & 5.1 & 1 && \enabled & \enabled & \disabled & \disabled & \enabled & \noprep & 286.6 & 4.7 & 239.8 \\ McRAPTOR&\showsource{~\cite{raptorjournal}} & \instLondon & 5.1 & 1 && \enabled & \enabled & \disabled & \disabled & \enabled & \noprep & 89.6 & 4.7 & 71.9 \\ \bottomrule \end{tabular} \end{table}

The table also contains the preprocessing time~(where applicable), the average number of label comparisons per stop, the average number of journeys computed by the algorithm, and its running time in milliseconds. Note that the number of journeys can be below 1 because some stops are unreachable for certain late departure times. References indicate the publications from which the figures are taken~(which may differ from the first publication); TED was run by the authors for this survey. (Our TED implementation uses a single-level bucket queue~\cite{d-aspft-69} and stops as soon as a vertex of the target stop has been extracted.) The columns labeled ``criteria'' indicate whether the algorithm minimizes arrival time~(arr), number of transfers~(tran), fare zones~(fare), reliability~(rel), and whether it computes range queries~(rng) over the full timetable period of 1, 2, or 7 days. Methods with multiple criteria compute Pareto sets.

Among algorithms without preprocessing, observe that those that do not use a graph~(RAPTOR and CSA) are consistently faster than their graph-based counterparts. Moreover, running Dijkstra on the time-expanded graph model~(TED) is significantly slower than on the time-dependent graph model~(TDD), since time-expanded graphs are much larger. For earliest arrival queries on metropolitan areas, CSA is the fastest algorithm without preprocessing, but preprocessing-based methods~(such as Transfer Patterns) can be even faster. For longer-range transit networks, preprocessing-based methods scale very well. CH takes 210 seconds to preprocess the long-distance train connections of Europe, while ACSA takes 8 hours to preprocess the full transit network of Germany. Transfer Patterns takes over 60 times longer to preprocess~(a full week of) the full transit network of Germany, but has considerably lower query times.

For multicriteria queries, RAPTOR is about an order of magnitude faster than Dijkstra-based approaches like LD and MLS. RAPTOR is twice as fast as TDD, while computing twice as many journeys on average. Adding further criteria~(such as fares and reliability) to MLS and RAPTOR increases the Pareto set, but performance is still reasonable for metropolitan-sized networks. Thanks to preprocessing, Transfer Patterns has the fastest queries overall, by more than an order of magnitude. Note that in public transit networks the optimization criteria are often positively correlated~(such as arrival time and number of transfers), which keeps the Pareto sets at a manageable size. Still, as the number of criteria increases, exact real-time queries become harder to achieve.

The reported figures for Transfer Patterns are based on preprocessing leveraging the frequency-based model with traffic days compression, which makes quadratic~(in the number of stops) preprocessing effort feasible. Consequently, hub stops and the three-leg heuristic are not required, and the algorithm is guaranteed to find the optimal solution. The data produced by the preprocessing is shown to be robust against large and area-wide delays, resulting in much less than 1\,\% of suboptimal journeys~\cite{bss-drtpp-13}~(not shown in the table).

For range queries, preprocessing-based techniques~(CH, ACSA, Transfer Patterns) scale better than CSA or SPCS. For full multicriteria range queries~(considering transfers), Transfer Patterns is by far the fastest method, thanks to preprocessing. Among search-based methods, CSA is faster than rRAPTOR by a factor of two, although it does twice the amount of work in terms of label comparisons. Note, however, that while CSA cannot scale to smaller time ranges by design~\cite{dpsw-isftr-13}, the performance of rRAPTOR depends linearly on the number of journeys departing within the time range~\cite{dpw-rbptr-12}. For example, for 2-hour range queries rRAPTOR computes~15.9~journeys taking only~61.3\,ms on average~\cite{raptorjournal}~(not reported in the table). Guidebook routes covering about~80\,\%~of the optimal results~(for the full period) can be computed in a fraction of a millisecond~\cite{bs-fbgr-14}.

\section{Multimodal Journey Planning} \label{sec:mm}

We now consider journey planning in a multimodal scenario. Here, the general problem is to compute journeys that \emph{reasonably} combine different modes of transportation by a \emph{holistic} algorithmic approach. That is, not only does an algorithm consider each mode of transportation in isolation, but it also optimizes the choice~(and sequence) of transportation modes in some integrated way. Transportation modes that are typically considered include~(unrestricted) walking,~(unrestricted) car travel,~(local and long-distance) public transit, flight networks, and rental bicycle schemes. We emphasize that our definition of ``multimodal'' requires some diversity from the transportation modes, i.\,e., both unrestricted and schedule-based variants should be considered by the algorithm. For example, journeys that only use buses, trams, or trains are not truly multimodal~(according to our definition), since these transportation modes can be represented as a single public transit schedule and dealt with by algorithms from Section~\ref{sec:public}.

In fact, considering modal transfers explicitly by the algorithm is crucial in practice, since the solutions it computes must be \emph{feasible}, excluding sequences of transportation modes that are impossible for the user to take~(such as a private car between train rides). Ideally, even user preferences should be respected. For example, some users may prefer taxis over public transit at certain parts of the journey, while others may not.

A general approach to obtain a multimodal network is to first build an individual graph for each transportation mode, then merge them into a single multimodal graph with \emph{link arcs}~(or vertices) added to enable modal transfers~\cite{p-mmrp-09,dpw-ammrp-09,yl-a-12}. Typical examples~\cite{p-mmrp-09,dpw-ammrp-09} model car travel and walking as time-independent~(static) graphs, public transit networks using the realistic time-dependent model~\cite{pswz-emtip-08}, and flight networks using a dedicated flight model~\cite{dpwz-erpfn-09}. Beyond that, Kirchler et al.~\cite{klpc-ualt-11,klc-alcas-12} compute multimodal journeys in which car travel is modeled as a time-dependent network in order to incorporate historic data on rush hours and traffic congestion~(see Section~\ref{ssec:extensions} for details).

\paragraph{Overview.} The remainder of this section discusses three different approaches to the multimodal problem. The first~(Section~\ref{sec:mm:cost}) considers a combined cost function of travel time with some penalties to account for modal transfers. The second approach~(Section~\ref{sec:mm:lcsp}) uses the label-constrained shortest path problem to obtain journeys that explicitly include~(or exclude) certain sequences of transportation modes. The final approach~(Section~\ref{sec:mm:mc}) computes Pareto sets of multimodal journeys using a carefully chosen set of optimization criteria that aims to provide diverse~(regarding the transportation modes) alternative journeys.

\subsection{Combining Costs} \label{sec:mm:cost}

To aim for journeys that reasonably combine different transport modes, one may use penalties in the objective function of the algorithm. These penalties are often considered as a linear combination with the primary optimization goal~(typically travel time). Examples for this approach include Aifadopoulou et al.~\cite{azc-mopap-07}, who present a linear program that computes multimodal journeys. The TRANSIT algorithm~\cite{aw-fmopm-12} also uses a linear utility function and incorporates travel time, ticket cost, and ``inconvenience'' of transfers. Finally, Modesti and Sciomachen~\cite{ms-a-98} consider a combined network of unrestricted walking, unrestricted car travel, and public transit, in which journeys are optimized according to a linear combination of several criteria, such as cost and travel time. Moreover, their utility function incorporates user preferences on the transportation modes.

\subsection{Label-Constrained Shortest Paths} \label{sec:mm:lcsp}

The \emph{label-constrained shortest paths}~\cite{bjm-flcpp-00} approach computes journeys that explicitly obey certain constraints on the modes of transportation. It defines an alphabet~$\alphabet$ of modes of transportation and labels each arc of the graph by the appropriate symbol from~$\alphabet$. Then, given a language~$\alanguage$ over~$\alphabet$ as additional input to the query, any journey~(path) must obey the constraints imposed by the language~$\alanguage$, i.\,e., the concatenation of the labels along the path must satisfy~$\alanguage$. The problem of computing \emph{shortest} label-constrained paths is tractable for \emph{regular} languages~\cite{bjm-flcpp-00}, which suffice to model reasonable transport mode constraints in multimodal journey planning~\cite{bbjkm-ccspp-02,bbhkmw-elcsp-08}. Even restricted classes of regular languages can be useful, such as those that impose a hierarchy of transport modes~\cite{bbm-o-06,p-mmrp-09,dpw-ammrp-09,klpc-ualt-11,klc-alcas-12,yl-a-12} or Kleene languages that can only globally exclude~(and include) certain transport modes~\cite{grst-rpfer-12}.

Barrett et al.~\cite{bjm-flcpp-00} have proven that the label-constrained shortest path problem is solvable in deterministic polynomial time. The corresponding algorithm, called \emph{label-constrained shortest path problem Dijkstra}~(LCSPP-D), first builds a product network~$\mmgraph$ of the input~(the multimodal graph) and the~(possibly nondeterministic) finite automaton that accepts the regular language~$\alanguage$. For given source and target vertices~$\sourcevertex,\targetvertex$~(referring to the original input), the algorithm determines origin and destination sets of product vertices from~$\mmgraph$, containing those product vertices that refer to~$s$/$t$ and an initial/final state of the automaton. Dijkstra's algorithm is then run on~$\mmgraph$ between these two sets of product vertices. In a follow-up experimental study, Barrett et al.~\cite{bbjkm-ccspp-02} evaluate this algorithm using linear regular languages, a special case.

Basic speedup techniques, such as bidirectional search~\cite{d-lpe-62}, \astar~\cite{hnr-afbhd-68}, and heuristic \astar~\cite{sv-speg-86} have been evaluated in the context of multimodal journey planning in~\cite{h-epssp-08} and~\cite{bbhkmw-elcsp-09}. Also, Pajor~\cite{p-mmrp-09} combines the LCSPP-D algorithm with time-dependent Dijkstra~\cite{ch-tsrtn-66} to compute multimodal journeys that contain a time-dependent subnetwork. He also adapts and analyzes bidirectional search~\cite{d-lpe-62}, ALT~\cite{gh-cspas-05}, Arc Flags~\cite{l-aeept-09,hkms-fppsp-09}, and shortcuts~\cite{v-ispat-78} with respect to LCSPP.

\paragraph{Access-Node Routing.} The \emph{Access-Node Routing}~(ANR)~\cite{dpw-ammrp-09} algorithm is a speedup technique for the label-constrained shortest path problem~(LCSPP). It handles \emph{hierarchical languages}, which allow constraints such as restricting walking and car travel to the beginning and end of the journey. It works similarly to Transit Node Routing~\cite{bfmss-itcsp-07,bfss-frrnt-07,bfm-uspqt-09,ss-racts-09} and precomputes for each vertex~$\vertex$ of the road~(walking and car) network its relevant set of entry~(and exit) points~(\emph{access nodes}) to the public transit and flight networks. More precisely, for any shortest path~$\apath$ originating from vertex~$\vertexa$~(of the road network) that also uses the public transit network, the first vertex~$\vertexb$ of the public transit network on~$\apath$ must be an access node of~$\vertex$. The query may skip over the road network by running a multi-source multi-target algorithm on the~(much smaller) transit network between the access nodes of~$\sourcevertex$ and~$\targetvertex$, returning the journey with earliest combined arrival time.

The \emph{Core-Based ANR}~\cite{dpw-ammrp-09} method further reduces preprocessing space and time by combining ANR with contraction. As in Core-ALT~\cite{bdsssw-chgds-10,dn-crdtd-12}, it precomputes access nodes only for road vertices in a much smaller core~(overlay) graph. The query algorithm first~(quickly) determines the relevant core vertices of~$\sourcevertex$ and~$\targetvertex$~(i.\,e., those covering the branches of the shortest path trees rooted at~$\sourcevertex$ and~$\targetvertex$), then runs a multi-source multi-target ANR query between them.

Access-Node Routing has been evaluated on multimodal networks of intercontinental size that include walking, car travel, public transit, and flights. Queries run in milliseconds, but preprocessing time strongly depends on the density of the public transit and flight networks~\cite{dpw-ammrp-09}. Moreover, since the regular language is used during preprocessing, it can no longer be specified at query time without loss of optimality.

\paragraph{State-Dependent ALT.} Another multimodal speedup technique for LCSPP is \emph{State-Dependent ALT}~(SDALT)~\cite{klpc-ualt-11}. It augments the ALT algorithm~\cite{gh-cspas-05} to overcome the fact that lower bounds from a vertex~$\vertex$ may depend strongly on the current state~$\astate$ of the automaton~(expressing the regular language) with which~$\vertex$ is scanned. SDALT thus uses the automaton to precompute state-dependent distances, providing lower bound values per vertex \emph{and} state. For even better query performance, SDALT can be extended to use more aggressive~(and potentially incorrect) bounds to guide the search toward the target, relying on a label-correcting algorithm~(which may scan vertices multiple times) to preserve correctness~\cite{klc-alcas-12}. SDALT has been evaluated~\cite{klpc-ualt-11,klc-alcas-12} on a realistic multimodal network covering the Île-de-France area~(containing Paris) incorporating rental and private bicycles, public transit, walking, and a time-dependent road network for car travel. The resulting speedups are close to 30. Note that SDALT, like ANR, also predetermines the regular language constraints during preprocessing.

\paragraph{Contraction Hierarchies.}

Finally, Dibbelt et al.~\cite{dpw-ucmmr-12} have adapted Contraction Hierarchies~\cite{gssv-erlrn-12} to LCSPP, handling arbitrary mode \emph{sequence} constraints. The resulting User-Constrained Contraction Hierarchies~(UCCH) algorithm works by~(independently) only contracting vertices whose incident arcs belong to the same modal subnetwork. All other vertices are kept uncontracted. The query algorithm runs in two phases. The first runs a regular CH query in the subnetworks given as initial or final transport modes of the sequence constraints until the uncontracted \emph{core graph} is reached. Between these entry and exit vertices, the second phase then runs a regular LCSPP-Dijkstra restricted to the~(much smaller) core graph. Query performance of UCCH is comparable to Access-Node Routing, but with significantly less preprocessing time and space. Also, in contrast to ANR, UCCH also handles arbitrary mode sequence constraints at query time.

\subsection{Multicriteria Optimization} \label{sec:mm:mc}

While label constraints are useful to define feasible journeys, computing the~(single) shortest label-constrained path has two important drawbacks. First, in order to define the constraints, users must know the characteristics of the particular transportation network; second, alternative journeys that combine the available transportation modes differently are not computed. To obtain a set of diverse alternatives, multicriteria optimization has been considered.

The criteria optimized by these methods usually include arrival time and, for each mode of transportation, some mode-dependent optimization criterion~\cite{ddpww-cmjp-13,bbs-rdmmr-13}. The resulting Pareto sets will thus contain journeys with different usage of the available transportation modes, from which users can choose their favorites.

Delling et al.~\cite{ddpww-cmjp-13} consider networks of metropolitan scale and use the following criteria as proxies for ``convenience'': number of transfers in public transit, walking duration for the pedestrian network, and monetary cost for taxis. They observe that simply applying the MLS algorithm~\cite{h-bpp-79,m-omspp-84,t-rmlw-95,m-vvpgm-99} to a comprehensive multimodal graph turns out to be slow, even when partial contraction is applied to the road and pedestrian networks, as in UCCH~\cite{dpw-ucmmr-12}. To get better query performance, they extend RAPTOR~\cite{dpw-rbptr-12} to the multimodal scenario, which results in the \emph{multimodal multicriteria RAPTOR} algorithm~(MCR)~\cite{ddpww-cmjp-13}. Like RAPTOR, MCR operates in rounds~(one per transfer) and computes Pareto sets of optimal journeys with exactly~$i$ transfers in round~$i$. It does so by running, in each round, a dedicated subalgorithm~(RAPTOR for public transit; MLS for walking and taxi) which obtains journeys with the respective transport mode as their last leg.

Since with increasing number of optimization criteria the resulting Pareto sets tend to get very large, Delling et al.~identify the most significant journeys in a quick postprocessing step by a scoring method based on fuzzy logic~\cite{z-fl-88}. For faster queries, MCR-based heuristics~(which relax domination during the algorithm) successfully find the most significant journeys while avoiding the computation of insignificant ones in the first place.

Bast et al.~\cite{bbs-rdmmr-13} use MLS with contraction to compute multimodal multicriteria journeys at a metropolitan scale. To identify the significant journeys of the Pareto set, they propose a method called \emph{Types aNd Thresholds}~(TNT). The method is based on a set of simple \emph{axioms} that summarize what most users would consider as unreasonable multimodal paths. For example, if one is willing to take the car for a large fraction of the trip, one might as well take it for the whole trip. Three types of reasonable trips are deduced from the axioms: (1)~only car, (2)~arbitrarily much transit and walking with no car, and (3)~arbitrarily much transit with little or no walking and car. With a concrete threshold for ``little''~(such as 10 minutes), the rules can then be applied to filter the reasonable journeys. As in~\cite{ddpww-cmjp-13}, filtering can be applied during the algorithm to prune the search space and reduce query time. The resulting sets are fairly robust with respect to the choice of threshold.

\section{Final Remarks} \label{sec:conclusion}

The last decade has seen astonishing progress in the performance of shortest path algorithms on transportation networks. For routing in road networks, in particular, modern algorithms can be up to seven orders of magnitude faster than standard solutions. Successful approaches exploit different properties of road networks that make them easier to deal with than general graphs, such as goal direction, a strong hierarchical structure, and the existence of small separators. Although some early acceleration techniques relied heavily on geometry~(road networks are after all embedded on the surface of the Earth), no current state-of-the-art algorithm makes explicit use of vertex coordinates~(see Table~\ref{tab:road_simple}). While one still sees the occasional development~(and publication) of geometry-based algorithms they are consistently dominated by established techniques. In particular, the recent Arterial Hierarchies~\cite{zmxltz-spdqr-13} algorithm is compared to CH~(which has slightly slower queries), but not to other previously published techniques~(such as CHASE, HL, and TNR) that would easily dominate it. This shows that results in this rapidly-evolving area are often slow to reach some communities; we hope this survey will help improve this state of affairs.

Note that experiments on real data are very important, as properties of production data are not always accurately captured by simplified models and folklore assumptions. For example, the common belief that an algorithm can be augmented to include turn penalties without significant loss in performance turned out to be wrong for CH~\cite{dgpw-crp-11}.

Another important lesson from recent developments is that careful engineering is essential to unleash the full computational power of modern computer architectures. Algorithms such as CRP, CSA, HL, PHAST, and RAPTOR, for example, achieve much of their good performance by carefully exploiting locality of reference and parallelism~(at the level of instructions, cores, and even GPUs).

The ultimate validation of several of the approaches described here is that they have found their way into systems that serve millions of users every day. Several authors of papers cited in this survey have worked on routing-related projects for companies like Apple, Esri, Google, MapBox, Microsoft, Nokia, PTV, TeleNav, TomTom, and Yandex. Although companies tend to be secretive about the actual algorithms they use, in some cases this is public knowledge. TomTom uses a variant of Arc Flags with shortcuts to perform time-dependent queries~\cite{s-ttnh-12}. Microsoft's Bing Maps\footnote{\url{http://www.bing.com/blogs/site_blogs/b/maps/archive/2012/01/05/bing-maps-new-routing-engine.aspx}} use CRP for routing in road networks. OSRM~\cite{lv-rtros-11}, a popular route planning engine using OpenStreetMap data, uses CH for queries. The Transfer Patterns~\cite{bceghrv-frvlp-10} algorithm has been in use for public-transit journey planning on Google Maps\footnote{\url{http://www.google.com/transit}} since 2010. RAPTOR is currently in use by OpenTripPlanner\footnote{\url{http://opentripplanner.com}}.

These recent successes do not mean that all problems in this area are solved. The ultimate goal, a worldwide multimodal journey planner, has not yet been reached. Systems like Rome2Rio\footnote{\url{http://www.rome2rio.com}} provide a simplified first step, but a more useful system would take into account real-time traffic and transit information, historic patterns, schedule constraints, and monetary costs. Moreover, all these elements should be combined in a personalized manner. Solving such a general problem efficiently seems beyond the reach of current algorithms. Given the recent pace of progress, however, a solution may be closer than expected.

\clearpage

\bibliographystyle{plain} \begin{small} 

\end{small}

\end{document}